\documentclass[12pt,a4paper]{article}
\usepackage{amsfonts}
\usepackage{amsmath}
\usepackage{amssymb}
\usepackage{latexsym}
\usepackage{color}
\usepackage{subfig}
\usepackage[section]{placeins} 

\usepackage[dvips]{graphicx}

\numberwithin{equation}{section}

\newcommand{\x}{\mathsf{x}}
\newcommand{\X}{\mathsf{X}}
\newcommand{\BbbR}{\mathbb{R}}
\newcommand{\BbbZ}{\mathbb{Z}}

\newcommand{\Etilde}{\tilde{E}}

\DeclareMathOperator{\Realpart}{Re}
\DeclareMathOperator{\Imagpart}{Im}

\DeclareMathOperator{\arctanh}{arctanh}

\DeclareMathOperator{\arcsinh}{arcsinh}

\DeclareMathOperator{\sgn}{sgn}

\title{Static, stationary and inertial Unruh-DeWitt detectors on the BTZ black hole}

\author{Lee Hodgkinson${}^{1}$\thanks{pmxlh1@nottingham.ac.uk}
\ and
Jorma Louko${}^{1,2}$\thanks{jorma.louko@nottingham.ac.uk}
\\
\noalign{\vspace{3ex}}
\small{\it ${}^{1}$ School of Mathematical Sciences,
University of Nottingham,}\\
\small{\it Nottingham NG7 2RD, UK}
\\[1ex]
\small{\it ${}^{2}$ Kavli Institute for Theoretical Physics, 
University of California,}\\
\small{\it Santa Barbara, CA 93106-4030, USA}
\\[3ex]
\small{(June 2012; Revised August 2012)}
\\[3ex]
\small{Published in Phys.\ Rev.\ D 
\textbf{86}, 064031 (2012)}
}

\date{}

\begin{document}

\maketitle

\begin{abstract}

We examine an Unruh-DeWitt particle detector coupled to a scalar field
in three-dimensional curved spacetime. We first obtain a
regulator-free expression for the transition probability in an
arbitrary Hadamard state, working within first-order perturbation
theory and assuming smooth switching, and we show that both the transition 
probability and the instantaneous transition rate remain well defined
in the sharp switching limit. We then analyse a detector coupled to 
a massless conformally coupled field in the Hartle-Hawking vacua on the 
Ba\~nados-Teitelboim-Zanelli black hole, 
under both transparent and reflective boundary
conditions at the infinity. 
A~selection of stationary and freely-falling detector
trajectories are examined, including the co-rotating trajectories, for
which the response is shown to be thermal. Analytic results in a
number of asymptotic regimes, including those of large and small mass,
are complemented by numerical results in the interpolating regimes. The
boundary condition at infinity is seen to have a significant effect on
the transition rate.

\end{abstract}

\newpage

\section{Introduction}
\label{sec:intro}

The Unruh-DeWitt model for a particle detector \cite{unruh,deWitt} is
an important tool for probing the physics of quantum fields wherever
noninertial observers or curved backgrounds are present. In such cases
there is often no distinguished notion of a ``particle,'' analogous to
the plane-wave modes in Minkowski space, but an operational meaning
can be attached to the particle concept by analysing the transitions 
between the energy levels of a detector coupled to the field: upwards 
(respectively downwards) 
transitions can be interpreted as due to absorption (emission) 
of field quanta, or particles. The best-known applications of
this procedure are those for which the spectrum of transitions is
thermal, which is the case for uniformly linearly accelerated detectors in
Minkowski space~\cite{unruh,deWitt,takagi,Crispino:2007eb}, 
detectors at rest in the exterior
Schwarzschild black hole spacetime~\cite{hawking,Hartle:1976tp,Israel:1976ur}, 
and inertial detectors in de Sitter space~\cite{gibb-haw:dS}. 

With the Unruh-DeWitt detector, the fundamental quantity of interest is the probability of a transition between the energy eigenstates. In the framework of first order perturbation theory this probability is proportional to the response function, given by a Fourier transform of the Wightman distribution of the quantum field over the detector's worldline, weighted by a switching function that specifies how the interaction is turned on and off~\cite{byd,wald-smallbook}. 
The response function is mathematically well defined provided the state of the quantum field is regular in the Hadamard sense \cite{kay-wald,Decanini:2005gt} and the detector is switched on and off smoothly~\cite{Fewster:1999gj,junker,hormander-vol1,hormander-paper1}. Physically, the response function then gives the probability for the detector to have completed a quantum jump by the time the interaction with the field has ceased. 

A related quantity of interest is the transition rate, 
which can be defined as the derivative of the transition probability 
with respect to the total detection time and is observationally 
meaningful in terms of consequent measurements in identical 
ensembles of detectors~\cite{satz-louko:curved}. 
There are technical subtleties in isolating in the transition rate 
the effects that are 
merely due to the switch-on and switch-off 
and the effects that are genuinely due to the acceleration and 
to the quantum state of the 
field~\cite{Sriramkumar:1994pb,schlicht,Schlicht:thesis,Langlois,Langlois-thesis,louko-satz:profile}; 
however, a satisfactory treatment within first order perturbation theory is to start with 
a smoothly-switched detector and take a controlled sharp switching limit~\cite{satz-louko:curved,satz:smooth,Hodgkinson:2011pc}. 
In particular, 
in three-dimensional Minkowski space, with a massless scalar field in the Minkowski vacuum, this procedure yields a finite result both for the transition probability and the transition rate even in the sharp switching limit~\cite{Hodgkinson:2011pc}. 

In this paper we consider a detector in three-dimensional curved spacetime. 
In the first part of the paper we investigate a detector that is coupled to a scalar field in an arbitrary Hadamard state in an arbitrary spacetime. 
We give for the transition probability an expression that involves no distributional integrals, and we show that both the transition probability and the transition rate remain finite in the sharp switching limit. In the special case of a massless scalar field in the 
Minkowski vacuum of Minkowski spacetime, 
we recover the results of~\cite{Hodgkinson:2011pc}. 

In the second part of the paper we apply the transition rate formula
to a detector in the $(2+1)$-dimensional Ba\~nados-Teitelboim-Zanelli
(BTZ) black hole spacetime~\cite{BTZ,HBTZ}, for a massless conformally
coupled scalar field in a Hartle-Hawking vacuum state
\cite{Hartle:1976tp,Israel:1976ur,Lifschytz:1993eb,Carlip:1995} 
with transparent, Dirichlet or Neumann 
boundary conditions at the asymptotically anti-de~Sitter infinity.  
We first consider a
stationary detector outside the hole, switched on in the asymptotic
past. When the detector is co-rotating with the black hole horizon,
we verify that the transition rate is thermal 
in the co-rotating local Hawking temperature, 
in the sense of the
Kubo-Martin-Schwinger (KMS) property~\cite{Kubo:1957mj,Martin:1959jp}, 
as is to be expected from general properties of the Hartle-Hawking state 
\cite{Hartle:1976tp,Israel:1976ur,Lifschytz:1993eb,Gibbons:1976ue} 
and also from global embedding space (GEMS) considerations 
\cite{Deser:1997ri,Deser:1998bb,Deser:1998xb,Russo:2008gb}. 
We obtain analytic
results in a number of asymptotic regimes, including those of large
and small mass, and we provide numerical results in the interpolating
regimes. A~static detector outside a nonrotating black
hole is included as a special case. When the detector is stationary
but its angular velocity differs from that of the horizon, we find
that the transition rate breaks the KMS property already to quadratic 
order in the difference of the angular velocities of the
horizon and the detector.

We also consider a detector that falls into a nonrotating BTZ hole along a
radial geodesic. We analyse the time evolution of the 
transition rate, obtaining analytic results for
large black hole mass and large detector energy gap and numerical
results in other regimes. We find no evidence of thermality in the
transition rate, not even near the moment of maximum distance from the
black hole, and we trace this phenomenon to the non-thermal response
of an inertial detector in the corresponding vacuum state in pure
anti-de~Sitter space, as predicted from GEMS 
considerations~\cite{Deser:1997ri,Deser:1998bb,Deser:1998xb,Russo:2008gb}. 

We find that the boundary condition at the infinity has a 
significant effect on the transition rate for all of our detector trajectories. 
Typically, the 
Dirichlet condition gives a transition rate that varies least rapidly as a 
function of the detector's energy gap and the total detection time, 
owing to partial cancellations between terms in the 
Dirichlet Wightman function. Similar cancellations do not occur for the 
Neumann or transparent boundary condition. 

We begin in Section \ref{sec:detectormodels} with a brief review of the Unruh-DeWitt detector model, recalling in particular how the distributional character of the Wightman function needs to be addressed prior to taking a sharp switching limit. 
In Section \ref{sec:resp3d} we consider an arbitrary Hadamard state in an arbitrary three-dimensional 
spacetime, writing the transition probability without distributional integrals 
and obtaining the formulas for the transition probability and the transition rate in the sharp switching limit. 
Section \ref{sec:BTZoverview} briefly reviews relevant properties of the  
the BTZ black hole and the Hartle-Hawking vacua for a massless conformally coupled scalar field. A~stationary detector is considered in Section \ref{sec:corot} and a freely-falling detector in Section~\ref{sec:inertial}. 
Section \ref{sec:conclusions} presents concluding remarks. 
Technical steps in contour integral analyses and asymptotic expansions are delegated to six appendices. 

Our metric signature is $({-}{+}{+})$, and we use units in which 
$c=\hbar=1$. 
Spacetime points are denoted by sans-serif letters.
$O(x)$ denotes a quantity for which $O(x)/x$ is bounded as
$x\to0$, 
$o(x)$ denotes a quantity for which $o(x)/x \to0$ as
$x\to0$, 
$O(1)$ denotes a quantity that is bounded in the limit under
consideration, 
and 
$o(1)$ denotes a quantity that vanishes in the limit under
consideration.

\section{Unruh-DeWitt detector}
\label{sec:detectormodels}

In this section we briefly review the 
Unruh-DeWitt detector coupled to a scalar field, treated in 
first-order perturbation theory in the coupling~\cite{unruh,deWitt}. 

The detector is an idealised atom with two energy levels, denoted by 
$|0\rangle_d$ and~$|E\rangle_d$, 
with the respective energy eigenvalues $0$ and~$E$, where $E$ may be positive or negative. 
The detector is spatially pointlike and moves on a timelike worldline~$\x(\tau)$, 
parametrised by the detector's proper time $\tau$, in a spacetime of dimension two or higher. 
The detector interacts with a free real scalar field~$\phi$, 
of arbitrary mass and curvature coupling, by the interaction Hamiltonian 
\begin{equation}
H_{\text{int}}=c\chi(\tau)\mu(\tau)\phi\bigl(\x(\tau)\bigr) 
\ , 
\end{equation}
where $c$ is a small coupling constant and $\mu(\tau)$ is 
the atom's monopole moment operator. 
$\chi$~is the switching function, positive during the 
interaction and vanishing elsewhere; this function specifies 
how the detector is switched on and off. 
We assume $\chi$ to be smooth and of compact support, 
and we assume the trajectory to be smooth on the support of~$\chi$. 

Before the interaction begins, we assume the detector to be in the state $|0\rangle_d$ 
and the field to be in a state~$|\Psi\rangle$, and we assume $|\Psi\rangle$ to be regular in the sense that it satisfies the Hadamard property~\cite{Decanini:2005gt}. 
The detector-field system is hence initially in the composite state $|0\rangle_d\otimes|\Psi\rangle$. We are interested in the probability for the detector to be found in the state $|E\rangle_d$ after the interaction has ceased, regardless of the final state of the field. Working in first order perturbation theory, this probability factorises as~\cite{byd,wald-smallbook}
\begin{equation}
\label{eq:prob}
P(E)=c^2{|_d\langle0|\mu(0)|E\rangle_d|}^2\mathcal{F}\left(E\right)
\ , 
\end{equation}
where the response function $\mathcal{F}(E)$ 
encodes the information about the detector's trajectory, 
the initial state of the field and the way the detector has been switched on and off.
$\mathcal{F}(E)$~can be expressed as~\cite{schlicht}
\begin{equation}
\label{eq:respfunc}
\mathcal{F}(E)=2\,
\lim_{\epsilon\to0_+} 
\Realpart \int_{-\infty}^{\infty}\mathrm{d}u \,
\chi(u)\int_{0}^{\infty}\mathrm{d}s \,\chi(u-s)\, 
\mathrm{e}^{-iE s}\, W_\epsilon(u,u-s)\,,
\end{equation}
where $W_\epsilon(u,u-s)$ is a one-parameter family of functions that converge to the pull-back of the Wightman distribution to the detector's worldline~\cite{kay-wald,Fewster:1999gj,junker,satz-louko:curved}. The factors in front of $\mathcal{F}(E)$ in~\eqref{eq:prob} depend only on the internal structure of the detector and we  shall from now on drop them, referring to $\mathcal{F}(E)$ as the transition probability.

In summary, the response function $\mathcal{F}(E)$ answers the question 
``What is the probability of the detector 
to be observed in the state $|E\rangle_d$ after the 
interaction has ceased?". 
Given $\mathcal{F}(E)$, we may define the 
detector's \emph{transition rate} as the derivative of $\mathcal{F}(E)$ 
with respect to the total detection time~\cite{schlicht,louko-satz:profile}. Both $\mathcal{F}(E)$ and the transition rate depend not only on the initial state of the field and the detector's trajectory but also on the switching, and their behaviour in the sharp switching limit depends sensitively on the spacetime dimension~\cite{Hodgkinson:2011pc}.

Working directly with the expression \eqref{eq:respfunc} for $\mathcal{F}(E)$ 
is cumbersome because the limit $\epsilon\to0_+$ may not necessarily be taken pointwise under the 
integral~\cite{satz-louko:curved,schlicht,Schlicht:thesis,Langlois,Langlois-thesis,louko-satz:profile,satz:smooth,Hodgkinson:2011pc}. In Section \ref{sec:resp3d} we shall 
address this issue in three spacetime dimensions.

\section{Transition probability and transition rate in three spacetime dimensions}
\label{sec:resp3d}

In this section we specialise to three spacetime dimensions. We first rewrite the 
response function \eqref{eq:respfunc} in a form in which the regulator $\epsilon$ does not appear. 
We then take the sharp switching limit and show that both the transition probability and the transition rate remain well defined in this limit. 
We follow closely the procedure developed in~\cite{satz-louko:curved,louko-satz:profile,satz:smooth,Hodgkinson:2011pc}.

\subsection{Hadamard form of $W_{\epsilon}$}
\label{sec:resp3d:HadForm}

In a three-dimensional spacetime, the Wightman distribution $W(\x,\x')$ of a real scalar field in a Hadamard state can be represented by a family of functions with the short distance
form~\cite{Decanini:2005gt}
\begin{equation}
\label{eq:3dHadamard}
W_{\epsilon}(\x,\x')=\frac{1}{4\pi}\left[\frac{U(\x,\x')}{\sqrt{\tilde\sigma_{\epsilon}(\x,\x')}}
+\frac{H(\x,\x')}{\sqrt{2}}\right] , 
\end{equation}
where $\epsilon$ is a positive parameter, 
$\tilde\sigma(\x,\x')$ is the squared geodesic distance between $\x$ and~$\x'$, 
$\tilde\sigma_{\epsilon}(\x,\x'):=\tilde\sigma(\x,\x')+2i\epsilon\left[T(\x)-T(\x')\right]+\epsilon^2$ 
and $T$ is any globally-defined future-increasing $C^{\infty}$ function. 
The branch of the square root is such that the $\epsilon\to 0_+$ 
limit of the square root is positive when 
$\tilde\sigma(\x,\x')>0$~\cite{kay-wald,Decanini:2005gt}. 
Here $U(\x,\x')$ and $H(\x,\x')$ are symmetric biscalars that 
possess expansions of the form
\begin{subequations}
\begin{equation}
\label{eq:Udef}
U(\x,\x')=\sum^{\infty}_{n=0} U_n(\x,\x')\sigma^n(\x,\x'),
\end{equation}
\begin{equation}
\label{eq:Hdef}
H(\x,\x')=\sum^{\infty}_{n=0} H_n(\x,\x')\sigma^n(\x,\x'),
\end{equation}
\end{subequations}
where the coefficients $U_n(\x,\x')$ satisfy the recursion relations
\begin{eqnarray}
& & (n+1)(2n+1) U_{n+1} + (2n+1) U_{n+1 ; \mu} \sigma ^{; \mu}
-(2n+1) U_{n+1} \Delta ^{-1/2}{\Delta^{1/2}}_{;\mu} \sigma^{; \mu}
\nonumber \\
&  & + \left( \Box_x -m^2 -\xi R \right) U_n =0 , 
\quad n=0,1,2,\ldots,
\label{eq:Urec}
\end{eqnarray}
with the boundary condition
\begin{equation}
\label{eq:U0}
U_0= \Delta^{1/2},
\end{equation}
and the coefficients $H_n(\x,\x')$ satisfy the recursion relations
\begin{eqnarray}
& & (n+1)(2n+3) H_{n+1} + 2(n+1) H_{n+1 ; \mu} \sigma ^{; \mu}- 
2(n+1) H_{n+1} \Delta ^{-1/2}{\Delta ^{1/2}}_{;\mu}\sigma^{; \mu}\nonumber \\
&  &  + \left( \Box_x-m^2-\xi R \right) H_n =0, 
\quad n=0,1,2,\ldots,
\label{eq:Hrec}
\end{eqnarray}
where $\sigma = \tfrac12 \tilde\sigma$, 
$\Delta(\x,\x')$ is the Van Vleck determinant, 
$m$ is the mass and $\xi$ is the curvature coupling parameter~\cite{Decanini:2005gt}.

The $i\epsilon$ prescription in \eqref{eq:3dHadamard} 
defines the singular part of $W(\x,\x')$: the action of the 
Wightman distribution is obtained by integrating $W_{\epsilon}(\x,\x')$ 
against test functions and taking the limit 
$\epsilon\to 0_+$ as in~\eqref{eq:respfunc}. 
This limit can be shown to be independent of the choice of 
global time function $T$~\cite{Fewster:1999gj,junker,hormander-vol1,hormander-paper1}.

\subsection{Transition probability without $i\epsilon$ regulator}

To evaluate the $\epsilon\to 0_+$ limit in~\eqref{eq:respfunc}, the main issue is at $s=0$,
where the Hadamard expansion~\eqref{eq:3dHadamard} shows that the integrand develops a
nonintegrable singularity as $\epsilon\to 0_+$. We shall work under the
assumption that any other singularities that the integrand develops as
$\epsilon\to 0_+$ are integrable. This will be the case in our applications
in Sections \ref{sec:corot} and~\ref{sec:inertial}. We note in passing that similar integrable
singularities can occur in any spacetime dimension, and the
four-dimensional results in~\cite{satz-louko:curved} should hence be understood to
involve a similar assumption.


We first split the $s$-integral in \eqref{eq:respfunc} 
into the subintervals $(0,\eta)$ and $(\eta,\infty)$ where $\eta=\sqrt{\epsilon}$. 
We then find the small $\epsilon$ expansions of each integral 
and finally combine the results. 


Let $I_{>}$ denote the $s\in(\eta,\infty)$ portion of the 
$s$-integral in~\eqref{eq:respfunc}, 
including the taking of the real part. 
Let $W_0$ denote the pointwise limit of $W_{\epsilon}$ as $\epsilon\to0_+$. 
Replacing $W_{\epsilon}$ by $W_0$ creates in $I_{>}$ the error 
\begin{align}
& 2 \Realpart \int_\eta^{\infty} \mathrm{d}s \,\chi(u-s)\,
\mathrm{e}^{-iE s}\bigl[ W_{\epsilon}(u,u-s)-W_0(u,u-s)\bigr] 
\nonumber 
\\
& =\frac{1}{2\pi} \Realpart \int_\eta^{\infty} \mathrm{d}s
\,\chi(u-s)\,\mathrm{e}^{-iEs} \, U(u,u-s) 
\left[ \frac{1}{\sqrt{\tilde\sigma+2i\epsilon\Delta T+\epsilon^2}}+\frac{i}{\sqrt{-\tilde\sigma}}\right], 
\label{eq:resp3d:error}
\end{align}
where $\tilde\sigma$ is evaluated at the pair
$(\mathsf{x},\mathsf{x}')=\bigl(\mathsf{x}(u),\mathsf{x}(u-s)\bigr)$, 
$\Delta T:=T\bigl(\x(u)\bigr)-T\bigl(\x(u-s)\bigr)$ 
and we recall that $\tilde\sigma<0$ as the trajectory is timelike.
The square brackets in~\eqref{eq:resp3d:error} have the same form as in the 
Minkowski analysis in \cite{Hodgkinson:2011pc} and obey similar estimates. 
The function $U$ has the small $s$ expansion
\begin{equation}
\label{eq:Usmalls}
U(u,u-s) = 1 +O(s^2)
\end{equation}
by virtue of~\eqref{eq:Udef}, \eqref{eq:U0} 
and the expansions $\tilde\sigma = -s^2+O(s^4)$ and $\Delta = 1+O(s^2)$. 
These observations show that the contribution from 
\eqref{eq:resp3d:error} to $I_{>}$ is $O(\eta)$. We hence have 
\begin{align}
I_{>}&=
2\int^{\infty}_{0}\,\mathrm{d}s\,\chi(u-s) 
\Realpart \left[\mathrm{e}^{-iEs} \, W_0(u,u-s)  \right] \ +O\left(\eta\right) , 
\label{eq:resp3d:upper}
\end{align}
where we have extended the lower limit to $0$ at the expense of an error 
that is contained in the $O\left(\eta\right)$ term since 
\eqref{eq:3dHadamard} shows that taking the real part 
under the integral makes the integrand regular at $s=0$. 


Let then $I_{<}$ denote the $s\in(0,\eta)$ portion of the 
$s$-integral in~\eqref{eq:respfunc}, 
including the taking of the real part. 
We have 
\begin{align}
I_{<}&=2\Realpart\int^{\eta}_0\,\mathrm{d}s\,\chi(u-s) \, \mathrm{e}^{-iEs} \, 
W_{\epsilon}(u,u-s)
\nonumber 
\\
&=\frac{1}{2\pi}\Realpart\int^{\eta}_0\,\mathrm{d}s\,\chi(u-s) \, 
\mathrm{e}^{-iEs}
\left[\frac{U(u,u-s)}{\sqrt{\tilde\sigma+2i\epsilon\Delta T+\epsilon^2}}+\frac{H(u,u-s)}{\sqrt{2}}\right].
\end{align}
Since $H(u,u-s)$ is a regular function,  
its contribution to $I_{<}$ is~$O\left(\eta\right)$. 
The contribution of the term involving $U$ can be found using the estimates 
of \cite{Hodgkinson:2011pc} and the expansion~\eqref{eq:Usmalls}. 
We find
\begin{equation}
\label{eq:resp3d:lower}
I_{<}=\chi(u)/4+O\left(\eta\right).
\end{equation}

Combining \eqref{eq:resp3d:upper} and~\eqref{eq:resp3d:lower}, 
and noting that their error estimates hold uniformly in~$u$, 
it is immediate to take the $\epsilon\to0_+$ limit in~\eqref{eq:respfunc}. 
We find 
\begin{align}
\mathcal{F}(E)
=\frac{1}{4}\int^{\infty}_{-\infty}\,\mathrm{d}u\,\left[\chi(u)\right]^2
\ +2\int_{-\infty}^{\infty}\mathrm{d}u\,\chi(u)\,\int_{0}^{\infty}\mathrm{d}s\,\chi(u-s) 
\Realpart\left[\mathrm{e}^{-iEs} \, W_0(u,u-s)\right]. 
\label{eq:resp3d:sharp:prob}
\end{align}
Note that the integrals in \eqref{eq:resp3d:sharp:prob} 
are regular, at $s=0$ by the Hadamard short-distance behaviour of~$W_0$, 
and at $s>0$ by our assumptions about the the singularity structure of $W_0$ at 
timelike-separated points.

\subsection{Sharp switching limit and the transition rate}
\label{sec:resp3d:sharp}

Up to now we have assumed the switching function $\chi$ to be smooth. 
When $\chi$ approaches the characteristic function of the interval
$[\tau_0,\tau_0+\tau]$ in a sufficiently controlled fashion~\cite{satz-louko:curved,satz:smooth,Hodgkinson:2011pc}, 
the integrands in \eqref{eq:resp3d:sharp:prob} remain regular, 
and taking the sharp switching limit under the
integral can be justified by dominated convergence. 
The transition probability takes the form
\begin{align}
\label{eq:resp3d:sharp:SSprob}
\mathcal{F}_\tau (E)
= 
\frac{\Delta\tau}{4}
+2\int^{\tau}_{\tau_0}\,\mathrm{d}u\,\int^{u-\tau_0}_{0}\,\mathrm{d}s
\Realpart\left[\mathrm{e}^{-iEs} \, W_0(u,u-s)\right], 
\end{align}
where $\Delta \tau := \tau - \tau_0$
and the subscript $\tau$ is included as a 
reminder of the dependence on the switch-off moment. 
Differentiation with respect to $\tau$ shows that 
the transition rate is given by 
\begin{align}
\label{eq:resp3d:sharp:rate}
\dot{\mathcal{F}}_{\tau}\left(E\right)
=
\frac{1}{4}+2\int^{\Delta\tau}_{0}\,\mathrm{d}s
\Realpart\left[\mathrm{e}^{-iEs} \, W_0(\tau,\tau-s)\right]
\ . 
\end{align}

Note that both 
\eqref{eq:resp3d:sharp:SSprob}
and 
\eqref{eq:resp3d:sharp:rate}
are well defined under our assumptions, and in the 
special case of a massless scalar field in the Minkowski 
vacuum they reduce to what was found in~\cite{Hodgkinson:2011pc}. 
Spacetime curvature has hence not introduced new 
singularities in the sharp switching limit. 

Note also that the pre-integral term $\frac14$ in \eqref{eq:resp3d:sharp:rate}
would have been missed if the limit $\epsilon\to0_+$ had been taken 
na\"\i{}vely under the integral in~\eqref{eq:respfunc}. 
Yet this term is essential: it was observed in \cite{Hodgkinson:2011pc}
that without this term one would not recover the standard thermal response 
for a uniformly linearly accelerated detector in 
Minkowski vacuum~\cite{takagi,Ooguri:1985nv}, 
and we shall see in Section \ref{sec:corot} that without this term we would not 
recover thermality for a co-rotating detector in the BTZ spacetime. 

\section{Detector in the BTZ spacetime}
\label{sec:BTZoverview}

We now turn to a detector in the BTZ
black hole spacetime \cite{BTZ,HBTZ}, specialising to a massless
conformally coupled scalar field in the Hartle-Hawking vacuum 
with transparent or reflective boundary conditions. In this section we briefly recall
relevant properties of the spacetime and the Wightman
function. 
More detail can be found in the review
in~\cite{Carlip:1995}.

Recall first that three-dimensional 
anti-de~Sitter 
spacetime $\text{AdS}_3$ may be defined as the submanifold 
\begin{equation}
\label{eq:AdShyperb}
-\ell^2 = 
-T_1^2-T_2^2 + X_1^2+X_2^2  
\end{equation}
in $\mathbb{R}^{2,2}$ with coordinates $(T_1,T_2,X_1,X_2)$ and metric
\begin{equation}
\label{eq:R22metric}
dS^2= -dT_1^2-dT_2^2 + dX_1^2+dX_2^2, 
\end{equation}
where $\ell$ is a positive parameter of dimension length. 
The BTZ black hole is obtained as a quotient of an open region in 
$\text{AdS}_3$ under a discrete isometry group~$\simeq\BbbZ$.
Specialising to a nonextremal black hole, 
a set of coordinates that are adapted to the relevant isometries
and cover the exterior region of the black hole are the 
BTZ coordinates $(t,r,\phi)$, defined in 
$\text{AdS}_3$ by 
\begin{eqnarray}
\label{eq:BoyerLind}
X_1=\ell\sqrt{\alpha}
  \sinh\left(\frac{r_+}{\ell}\phi-\frac{r_-}{\ell^2}t\right) &,&\quad\!\!
X_2=\ell\sqrt{\alpha-1}
  \cosh\left(\frac{r_+}{\ell^2}t-\frac{r_-}{\ell}\phi\right) , \nonumber\\[1ex]
T_1=\ell\sqrt{\alpha}
   \cosh\left(\frac{r_+}{\ell}\phi-\frac{r_-}{\ell^2}t\right) &,&\quad\!
T_2=\ell\sqrt{\alpha-1}
  \sinh\left(\frac{r_+}{\ell^2}t-\frac{r_-}{\ell}\phi\right) , 
\end{eqnarray}
where
\begin{equation}
\label{eq:alpha}
\alpha(r)=\left(\frac{r^2-r_-^2}{r_+^2-r_-^2}\right)  
\end{equation}
and the parameters $r_\pm$ satisfy $|r_-| < r_+$. 
The coordinate ranges covering the black hole exterior are 
$r_+ < r < \infty$, $-\infty < t < \infty$ and $-\infty < \phi < \infty$, 
and the $\BbbZ$ quotient is realised as the identification
$(t,r,\phi) \sim (t,r,\phi+2\pi)$. The outer 
horizon is at $r\to r_+$, 
and the asymptotically $\text{AdS}_3$ infinity is at $r\to\infty$. 
The metric takes the form
\begin{equation}
\label{eq:BTZmetric}
ds^2 = -( N^\perp)^2dt^2 + f^{-2}dr^2
  + r^2\left( d\phi + N^\phi dt\right)^2
\end{equation}
with
\begin{equation}
\label{eq:lapseshift}
N^\perp = f
= \left( -M + \frac{r^2}{\ell^2} + \frac{J^2}{4r^2} \right)^{1/2} ,
\quad N^\phi = - \frac{J}{2r^2} ,
\end{equation}
where the mass $M$ and the angular momentum $J$ are given by
\begin{equation}
\label{eq:MandJ}
M=(r_+^2+r_-^2)/\ell^2, \quad J=2r_+ r_-/\ell,
\end{equation}
and they satisfy $|J| < M \ell$.

In a quantum state invariant under~$\partial_{\phi}$, the 
Wightman function on the black hole spacetime can be expressed as an image sum of the 
corresponding $\text{AdS}_3$ Wightman function. If $G_A(\x,\x')$ denotes the 
$\text{AdS}_3$ Wightman function, the BTZ Wightman function reads \cite{Carlip:1995}
\begin{equation}
\label{eq:BTZWightmanSum}
G_{\text{BTZ}}(\x,\x')=\sum_n\,G_A(\x,\Lambda^n \x')
\end{equation}
where $\Lambda \x'$ denotes the action on $\x'$ of the group element 
$(t,r,\phi)\mapsto (t,r,\phi+2\pi)$,  
and the notation suppresses the distinction between points 
on $\text{AdS}_3$ and points on the quotient spacetime. 
The scalar field is assumed untwisted so that no 
additional phase factors appear in~\eqref{eq:BTZWightmanSum}.

We consider a massless, conformally coupled field, and the family of 
$\text{AdS}_3$ Wightman functions \cite{Carlip:1995}
\begin{equation}
\label{eq:AdSWightman}
G^{(\zeta)}_{A}(\x,\x')=\frac{1}{4\pi} \! 
\left(\frac{1}{\sqrt{{\Delta\X}^2(\x,\x')}}
-\frac{\zeta}{\sqrt{{\Delta\X}^2(\x,\x')+4\ell^2}}\right),
\end{equation}
where the parameter $\zeta\in\{0,1,-1\}$ 
specifies whether the boundary condition at infinity is respectively 
transparent, Dirichlet or Neumann. 
Here $\Delta\X^2(\x,\x')$ 
is the squared geodesic distance between $\x$ and $\x'$ 
in the flat embedding spacetime $\mathbb{R}^{2,2}$, 
given by 
\begin{equation}
\label{eq:R22Interval}
{\Delta\X}^2 (\x,\x') :=
-{(T_1-T_1^{'})}^2 -{(T_2-T_2^{'})}^2 
+ {(X_1-X_1^{'})}^2 +{(X_2-X_2^{'})}^2 \ ,
\end{equation}
and we have momentarily suppressed 
the $i\epsilon$ prescription in~\eqref{eq:AdSWightman}. 

With
\eqref{eq:BTZWightmanSum}
and~\eqref{eq:AdSWightman}, 
the transition rate
\eqref{eq:resp3d:sharp:rate}
takes the form 
\begin{align}
\dot{\mathcal{F}}_{\tau}(E)
=
\frac{1}{4}+\frac{1}{2\pi\sqrt{2}}\sum_{n=-\infty}^{\infty}\int^{\Delta\tau/\ell}_{0}\,
\mathrm{d}\tilde{s} \Realpart
\left[\mathrm{e}^{-iE\ell \tilde{s}}
\left(\frac{1}{\sqrt{\Delta\tilde{\X}^2_n}}-\frac{\zeta}{\sqrt{\Delta\tilde{\X}^2_n+2}}\right)
\right] , 
\label{eq:BTZ:rate}
\end{align}
where we have 
introduced the dimensionless integration variable 
$\tilde{s}:=s/\ell$ 
and written 
\begin{align}
\Delta\tilde{\X}_n^2 & :={\Delta\X}^2\bigl(\x(\tau),\Lambda^n\x(\tau-\ell\tilde{s})\bigr)
/\bigl(2\ell^2\bigr)
\nonumber 
\\[1ex]
& \hspace{.8ex} = -1+\sqrt{\alpha(r)\alpha(r')}
\cosh\!\left[(r_+/\ell)\left(\phi-\phi'-2\pi n \right)-(r_-/\ell^2)\left(t-t'\right)\right]
\nonumber
\\[1ex]
&\hspace{3.5ex}
-\sqrt{\bigl(\alpha(r)-1\bigr)\bigl(\alpha(r')-1\bigr)}
\cosh\!\left[(r_+/\ell^2)\left(t-t'\right)-(r_-/\ell)\left(\phi-\phi'-2\pi n \right)\right] ,
\label{eq:DXN2}
\end{align}
where the unprimed coordinates are evaluated at $\x(\tau)$ 
and the primed coordinates 
at $\x(\tau-\ell\tilde{s})$. 

What remains is to specify the branches of the square roots in~\eqref{eq:BTZ:rate}. 
As $s$ extends to a global time function in the relevant part of $\text{AdS}_3$, 
the prescription \eqref{eq:3dHadamard} 
implies that the square roots in \eqref{eq:BTZ:rate} are positive 
when the arguments are positive, 
and the square roots are analytically continued to 
negative values of the arguments 
by giving $s$ a small negative imaginary part. 


\section{Co-rotating detector in BTZ}
\label{sec:corot}

In this section we investigate the transition rate of a 
detector that is in the exterior region of the BTZ black hole and 
co-rotating with the horizon. 
As the detector is stationary, we take the switch-on to be in the asymptotic past. 
When the black hole is spinless, the detector is static.

\subsection{Transition rate and the KMS property}
\label{sec:corot:rate}

The angular velocity of the horizon is 
given by \cite{BTZ,HBTZ,Carlip:1995}
\begin{equation}
\label{eq:corot:Omega}
\Omega_H
= r_-/(r_+ \ell) , 
\end{equation}
and it has an operational meaning as the value that 
$\mathrm{d}\phi/\mathrm{d}t$ takes on any 
timelike worldline that crosses the horizon. 
The worldline of a detector that is in the exterior region and 
rigidly co-rotating with the horizon reads 
\begin{equation}
\label{eq:corot:traj}
r= \text{constant}\ , \ \ 
t=
\frac{\ell r_+ \tau}{\sqrt{r^2-r_+^2}\sqrt{r_+^2-r_-^2}} \ , 
\ \ 
\phi= \frac{r_- \tau}{\sqrt{r^2-r_+^2}\sqrt{r_+^2-r_-^2}} \ , 
\end{equation}
where
the value of $r$ specifies the radial location and $\tau$ is the proper time. 
We have set the additive constants in $t$ and $\phi$ 
to zero without loss of generality. 

Substituting \eqref{eq:corot:traj} into \eqref{eq:DXN2} and taking the 
switch-on to be in the asymptotic past, 
the transition rate \eqref{eq:BTZ:rate} takes the form
\begin{align}
\dot{\mathcal{F}}
(E)
& = 
\frac{1}{4}+\frac{1}{4\pi\sqrt{\alpha(r)-1}}
\sum_{n=-\infty}^{\infty}
\int^{\infty}_{0}\,\mathrm{d}\tilde{s} 
\Realpart \left[\mathrm{e}^{-iE\ell \tilde{s}}
\left(\frac{1}{\sqrt{K_n-\sinh^2\bigl(\Xi \tilde{s}+n\pi r_-/\ell \bigr)}} \right. \right. 
\nonumber\\[1ex]
& \hspace{20ex}
\left. \left. 
-\frac{\zeta}{\sqrt{Q_n-\sinh^2 \bigl(\Xi \tilde{s}+n\pi r_-/\ell\bigr)}}\right)\right] ,
\label{eq:corot:rate}
\end{align}
where 
\begin{subequations}
\label{eq:corot:pars}
\begin{align}
\label{eq:corot:Kn}
K_n & := {\bigl(1-\alpha^{-1}\bigr)}^{-1}\sinh^2 \bigl( n \pi r_+/\ell \bigr) , 
\\
\label{eq:corot:Qn}
Q_n & := K_n + {(\alpha-1)}^{-1} , 
\\
\label{eq:corot:Xi}
\Xi & := {\left(2\sqrt{\alpha-1} \, \right)}^{-1} , 
\end{align} 
\end{subequations}
$\alpha$ is given by~\eqref{eq:alpha}, 
and we have dropped the subscript $\tau$ from $\dot{\mathcal{F}}$ 
as the situation is stationary and 
the transition rate is independent of~$\tau$. 
The square roots in \eqref{eq:corot:rate} 
are positive 
for positive values of the argument, 
and they are analytically continued to 
negative values of the argument 
by giving $\tilde{s}$ a small negative imaginary part. 
Note that the integrand in \eqref{eq:corot:rate} has 
singularities at $\tilde{s}>0$, 
at places where the quantity under a square root changes sign, 
but all of these singularities are integrable. 

We show in Appendix \ref{app:A} that \eqref{eq:corot:rate} can be written as 
\begin{align}
\dot{\mathcal{F}}(E)
&= \frac{\mathrm{e}^{-\beta E\ell /2}}{2\pi}\sum^{\infty}_{n=-\infty}
\,\cos\bigl(n\beta E r_-\bigr)
\times
\nonumber 
\\
&
\hspace{6ex}
\times 
\int^{\infty}_0\,\mathrm{d}y\,\cos\bigl(y\beta E\ell /\pi\bigr)\Bigg(\frac{1}{\sqrt{K_n+\cosh^2 \! y}}
-\frac{\zeta}{\sqrt{Q_n+\cosh^2 \! y}}\Bigg) , 
\label{eq:corot:rate.evald.alt}
\end{align}
or alternatively as 
\begin{align}
&
\dot{\mathcal{F}}(E)
= \frac{1}{2(\mathrm{e}^{\beta E\ell }+1)}
-\frac{\zeta \mathrm{e}^{-\beta E\ell /2}}{2\pi}\int^{\infty}_0\,
\mathrm{d}y\,\frac{\cos\bigl(y\beta E\ell /\pi\bigr)}{\sqrt{Q_0+\cosh^2 \! y}}
\nonumber 
\\[1ex]
&+\frac{\mathrm{e}^{-\beta E\ell /2}}{\pi}\sum^{\infty}_{n=1}\,\cos\bigl(n\beta E  r_-\bigr)
\int^{\infty}_0\,\mathrm{d}y\,\cos\bigl(y\beta E\ell /\pi\bigr)
\Bigg(\frac{1}{\sqrt{K_n+\cosh^2 \! y}}
-\frac{\zeta}{\sqrt{Q_n+\cosh^2 \! y}}\Bigg) , 
\label{eq:corot:rate.evald}
\end{align}
where 
\begin{align}
\beta:=2\pi \sqrt{\alpha-1}
\ . 
\label{eq:beta-def}
\end{align}
It is evident from \eqref{eq:corot:rate.evald.alt} or \eqref{eq:corot:rate.evald} 
that $\dot{\mathcal{F}}$ depends on $E$ only via the dimensionless 
combination $\ell\beta E$. 
It is further evident that 
$\dot{\mathcal{F}}$ has the 
KMS property \cite{Kubo:1957mj,Martin:1959jp} 
\begin{equation}
\label{eq:corot:KMS}
\dot{\mathcal{F}}(E) = \mathrm{e}^{-\ell\beta E}\dot{\mathcal{F}}(-E) . 
\end{equation}
The transition rate is hence thermal in the temperature~${(\ell\beta)}^{-1}$. 

It can be verified that 
${(\ell\beta)}^{-1} = (-g_{00})^{-1/2}T_0$, where $T_0=\kappa_0/(2\pi)$, 
$\kappa_0$ is the surface gravity of the black hole with respect to the 
horizon-generating Killing vector $\partial_t + \Omega_H \partial_\phi$, 
and $g_{00}$ is the time-time component of the metric in 
coordinates adapted to the co-rotating observers. 
This means that the temperature ${(\ell\beta)}^{-1}$ of the detector response 
is the local Hawking temperature, 
obtained by renormalising the conventional  
Hawking temperature $T_0$ by the 
Tolman redshift factor at the detector's location. 
This is the temperature one would have expected 
by general properties of the Hartle-Hawking state 
\cite{Hartle:1976tp,Israel:1976ur,Lifschytz:1993eb}, 
including the periodicity of an 
appropriately-defined imaginary time coordinate~\cite{Gibbons:1976ue}, 
and also by GEMS considerations 
\cite{Deser:1997ri,Deser:1998bb,Deser:1998xb,Russo:2008gb}. 

Note that the expressions 
\eqref{eq:corot:rate.evald.alt}
and 
\eqref{eq:corot:rate.evald}
contain both terms of~\eqref{eq:resp3d:sharp:rate}, 
as shown in Appendix~\ref{app:A}. 
The pre-integral term $\tfrac14$ in \eqref{eq:resp3d:sharp:rate} 
is hence essential for recovering thermality: in \eqref{eq:corot:rate.evald} 
it can be regarded as having been grouped in the term $\tfrac12{(\mathrm{e}^{\beta E\ell }+1)}^{-1}$, 
which gives the transition rate in pure $\text{AdS}_3$ with the transparent boundary condition. 
The superficial Fermi-Dirac appearance of this pure $\text{AdS}_3$ term is a general feature of 
linearly-coupled scalar fields in 
odd spacetime dimensions~\cite{takagi,Lifschytz:1993eb,Ooguri:1985nv,Sriramkumar:2002nt}.

\subsection{Asymptotic regimes}
\label{sec:corot:asymptotics}

We consider the behaviour of the transition rate 
\eqref{eq:corot:rate.evald} in three asymptotic regimes. 

First, suppose $r_+\to\infty$ so that $r_-/r_+$ and $r/r_+$ are fixed. 
Physically, this is the limit of a large black hole with fixed~$J/M$, 
and the detector is assumed not to be close to the 
black hole horizon. Note that 
$\alpha$ and $\beta$ remain fixed in this limit. 
It follows from 
\eqref{eq:alpha} and \eqref{eq:corot:pars}
that in \eqref{eq:corot:rate.evald} 
this is the limit in which $K_n$ and $Q_n$ with $n\ge1$ are large. 
Assuming that $E$ is fixed and nonzero, and using formula \eqref{eq:Jall3}
in Appendix~\ref{app:B}, we find 
\begin{align}
& 
\dot{\mathcal{F}}
(E)
= \frac{1}{2(\mathrm{e}^{\beta E\ell }+1)}
-\frac{\zeta \mathrm{e}^{-\beta E\ell /2}}{2\pi}\int^{\infty}_0\,
\mathrm{d}y\,\frac{\cos\bigl(y\beta E\ell /\pi\bigr)}{\sqrt{Q_0+\cosh^2 \! y}}
\nonumber 
\\[1ex]
&\hspace{1ex}
+\frac{\mathrm{e}^{-\beta E\ell /2}
\cos\bigl(\beta E r_- \bigr)}{\sqrt{\pi}\beta E\ell }
\times
\nonumber\\[1ex]
&\hspace{3ex}
\times 
\Bigg\{\Imagpart 
\Bigg[
\left(\frac{{(4K_1})^{i\beta E\ell /(2\pi)}}{\sqrt{K_1}}-
\frac{\zeta{(4Q_1})^{i\beta E\ell /(2\pi)}}{\sqrt{Q_1}}
\right)
\Gamma\left(1+\frac{i\beta E\ell }{2\pi}\right)
\Gamma\left(\frac12 - \frac{i\beta E\ell }{2\pi}\right)
\Bigg]
\nonumber\\[1ex]
&\hspace{8ex}
+O\left(\mathrm{e}^{-2\pi r_+/\ell}\right)\Bigg\}
\ , 
\label{eq:corot:larger+}
\end{align}
where the displayed next-to-leading term 
comes from the $n=1$ term in 
\eqref{eq:corot:rate.evald}
and
is of order~$\mathrm{e}^{-\pi r_+/\ell}$. 
The corresponding formula for $E=0$ can be obtained from formula \eqref{eq:Jall30} in 
Appendix \ref{app:B} and has a next-to-leading term 
of order $r_+\mathrm{e}^{-\pi r_+/\ell}$. 

Next, suppose that $r_+\to0$ so that $r_-/r_+$ and $r/r_+$ are again fixed. 
This is the limit of a small black hole. 
Note that 
$\alpha$ and $\beta$ are again fixed. 
The dominant behaviour comes now from the sum over $n$ 
and can be estimated by the Riemann sum technique of Appendix~\ref{app:C}. 
We find 
\begin{align}
\dot{\mathcal{F}}(E)
&=\frac{\ell \mathrm{e}^{-\beta E\ell /2}}{\pi^2r_+}
\int^{\infty}_0 \,\mathrm{d}v
\int^{\infty}_0\,\mathrm{d}y\,
\cos\!\left(\frac{v \beta E\ell  r_-}{\pi r_+}\right)
\cos\!\left(\frac{y\beta E\ell }{\pi}\right)
\times
\nonumber\\[1ex]
&\hspace{2ex} \times 
\left[ 
{\left(\frac{\alpha\sinh^2 \! v}{(\alpha-1)}+\cosh^2{y}\right)}^{\!-1/2}
-\zeta {\left(\frac{1 + \alpha \sinh^2 \! v}{(\alpha-1)}+\cosh^2 \! y \right)}^{\!-1/2}
\right]
+\frac{o(1)}{r_+}. 
\label{eq:corot:smallr+}
\end{align}
The leading term is proportional to $1/r_+$ and it hence diverges in the limit of a small black hole. 

Finally, suppose that $E \to \pm \infty$ with the other quantities fixed. 
The analysis of Appendix \ref{app:D} shows that each integral term in \eqref{eq:corot:rate.evald}
is oscillatory in~$E$, with an envelope that falls off as 
$1/\sqrt{-E}$ 
at $E \to -\infty$ but exponentially at $E \to +\infty$. 
Applying this estimate to the lowest few values of $n$ in \eqref{eq:corot:rate.evald} 
should be a good estimate to the whole sum when $r_+/\ell$ is large. 
We have not attempted to estimate the whole sum at $E \to \pm \infty$ when $r_+/\ell$ is small.

\subsection{Numerical results}
\label{sec:corot:results}

We now turn to numerical evaluation of the transition rate~\eqref{eq:corot:rate.evald}. 
We are particularly interested in the interpolation between 
the asymptotic regimes identified in subsection~\ref{sec:corot:asymptotics}. 

$\dot{\mathcal{F}}$ \eqref{eq:corot:rate.evald} 
depends on five independent variables. 
Two of these are the mass and the angular momentum of the black hole, 
encoded in the dimensionless parameters $r_+/\ell$ and $r_-/\ell$. 
The third is the location of the detector, entering 
$\dot{\mathcal{F}}$ only in the dimensionless 
combination $\alpha$~\eqref{eq:alpha}. 
The fourth is the detector's energy gap~$E$, 
entering $\dot{\mathcal{F}}$ only in the dimensionless combination $\beta E\ell$ 
where $\beta$ was given in~\eqref{eq:beta-def}. 
The last one is the discrete parameter $\zeta \in \{0,1,-1\}$ 
which specifies the boundary condition at infinity. 

We plot $\dot{\mathcal{F}}$ as a function of $\ell\beta E$, grouping the plots in 
triplets where $\zeta$ runs over its three values and the other three parameters are fixed. 
We proceed from large $r_+/\ell$ towards small~$r_+/\ell$. 

\begin{figure}[b!]  
  \centering
  \subfloat[$\zeta=0$]{\label{fig:Num_bc0_rp10_rm0_a4_and_a100_k3}\includegraphics[width=0.3\textwidth]{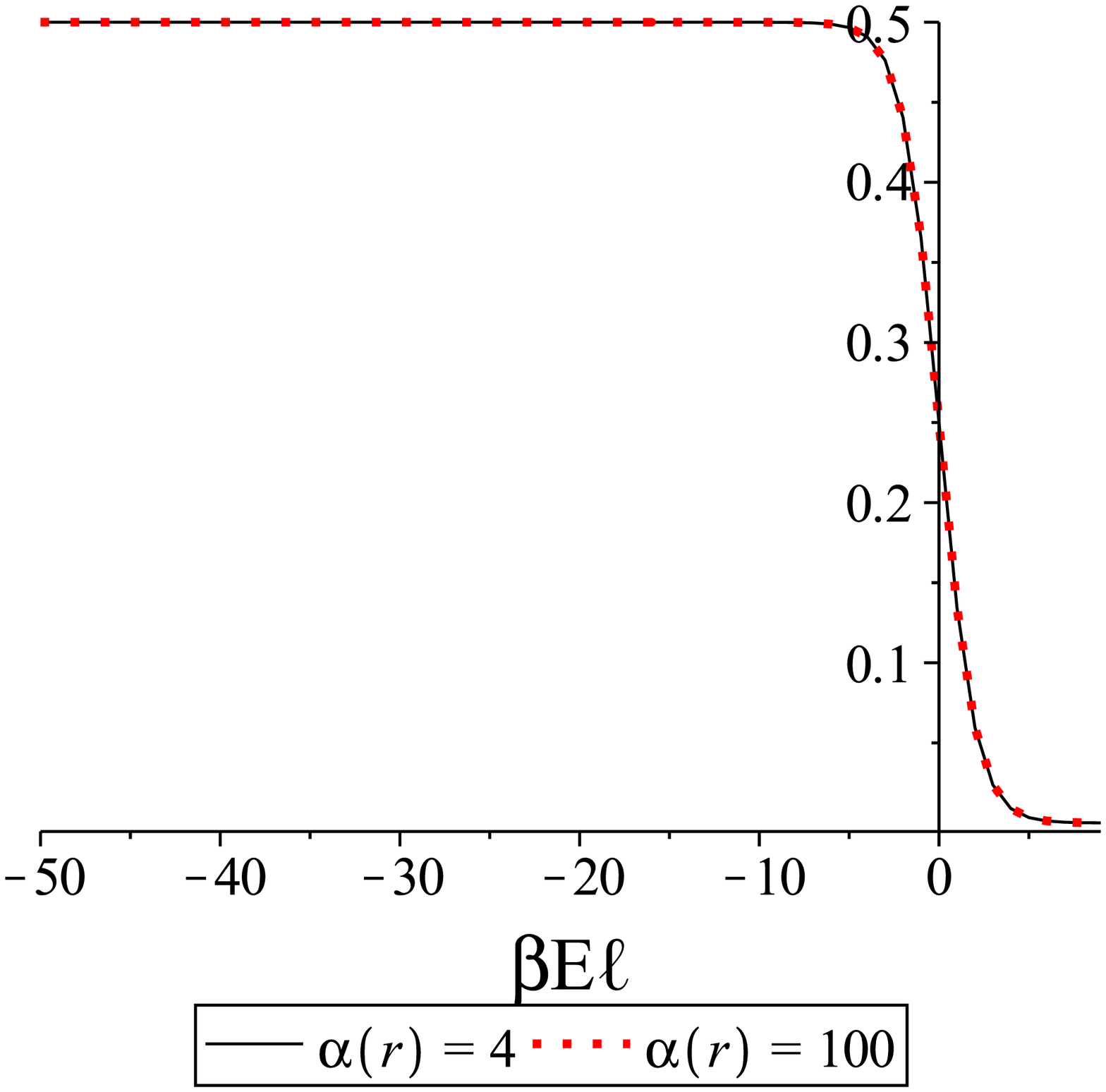}}            
  \subfloat[$\zeta=1$]{\label{fig:Num_bc1_rp10_rm0_a4_and_a100_k3}\includegraphics[width=0.3\textwidth]{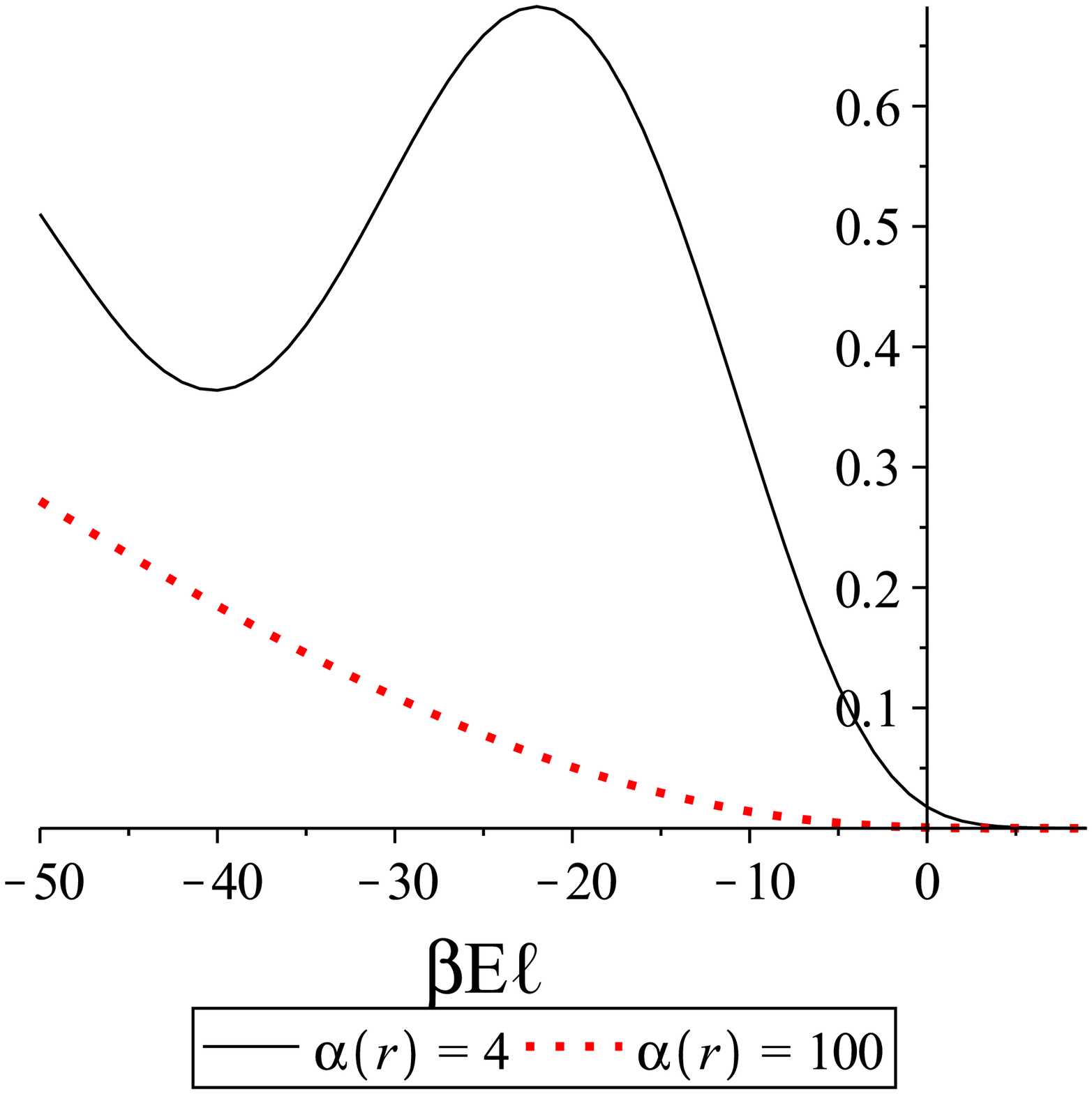}}  
  \subfloat[$\zeta=-1$]{\label{fig:Num_bc-1_rp10_rm0_a4_and_a100_k3}\includegraphics[width=0.3\textwidth]{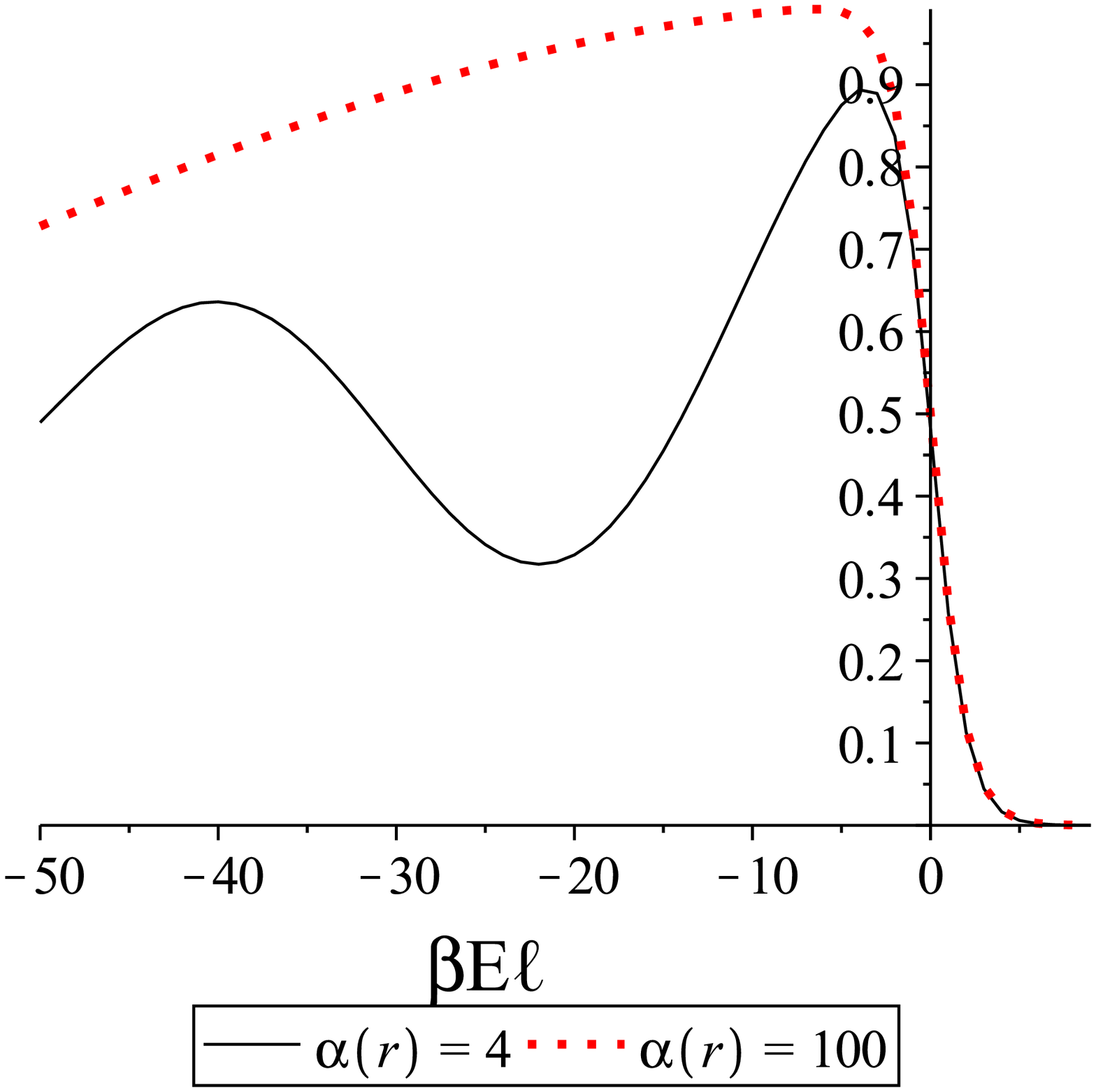}}  
\caption{$\dot{\mathcal{F}}$ as a function of $\beta E\ell$ for 
$r_+/\ell=10$ and $r_-/\ell=0$, with $\alpha=4$ (solid) and $\alpha=100$ (dotted). 
Numerical evaluation from \eqref{eq:corot:rate.evald} with $n\le3$.}
\label{fig:transrate_rp10}
\end{figure}

In the regime $r_+/\ell \gtrsim 3$, numerics confirms that 
the $n\ge1$ terms in \eqref{eq:corot:rate.evald} are small. 
$\dot{\mathcal{F}}$ therefore depends on 
$r_+/\ell$ and $r_-/\ell$ significantly only through~$\beta$, 
that is, through the local temperature. 
The detector's location enters 
$\dot{\mathcal{F}}$ in part via 
$\beta$ \eqref{eq:beta-def}, but also via $Q_0$ 
in \eqref{eq:corot:rate.evald}, 
and the latter affects only the the boundary 
conditions $\zeta=1$ and $\zeta=-1$, in opposite directions. 
Plots for $r_+/\ell =10$ are shown in Figure~\ref{fig:transrate_rp10}. 

\begin{figure}[t!]
  \centering
  \subfloat[$\zeta=0$]{\label{fig:Num_bc0_rp1_rm0_rm0pt99_a2_k3}\includegraphics[width=0.3\textwidth]{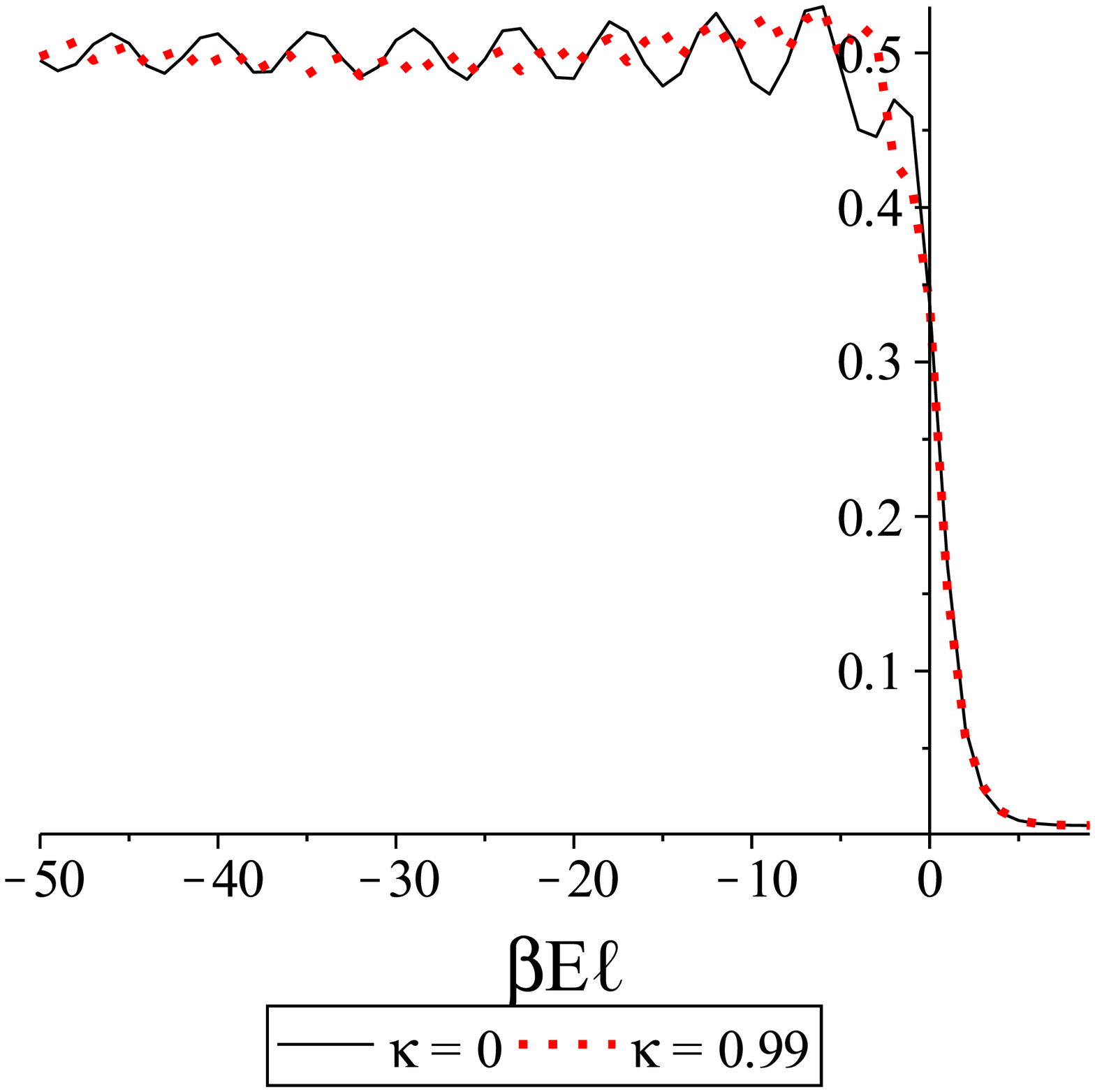}}            
  \subfloat[$\zeta=1$]{\label{fig:Num_bc1_rp1_rm0_rm0pt99_a2_k3}\includegraphics[width=0.3\textwidth]{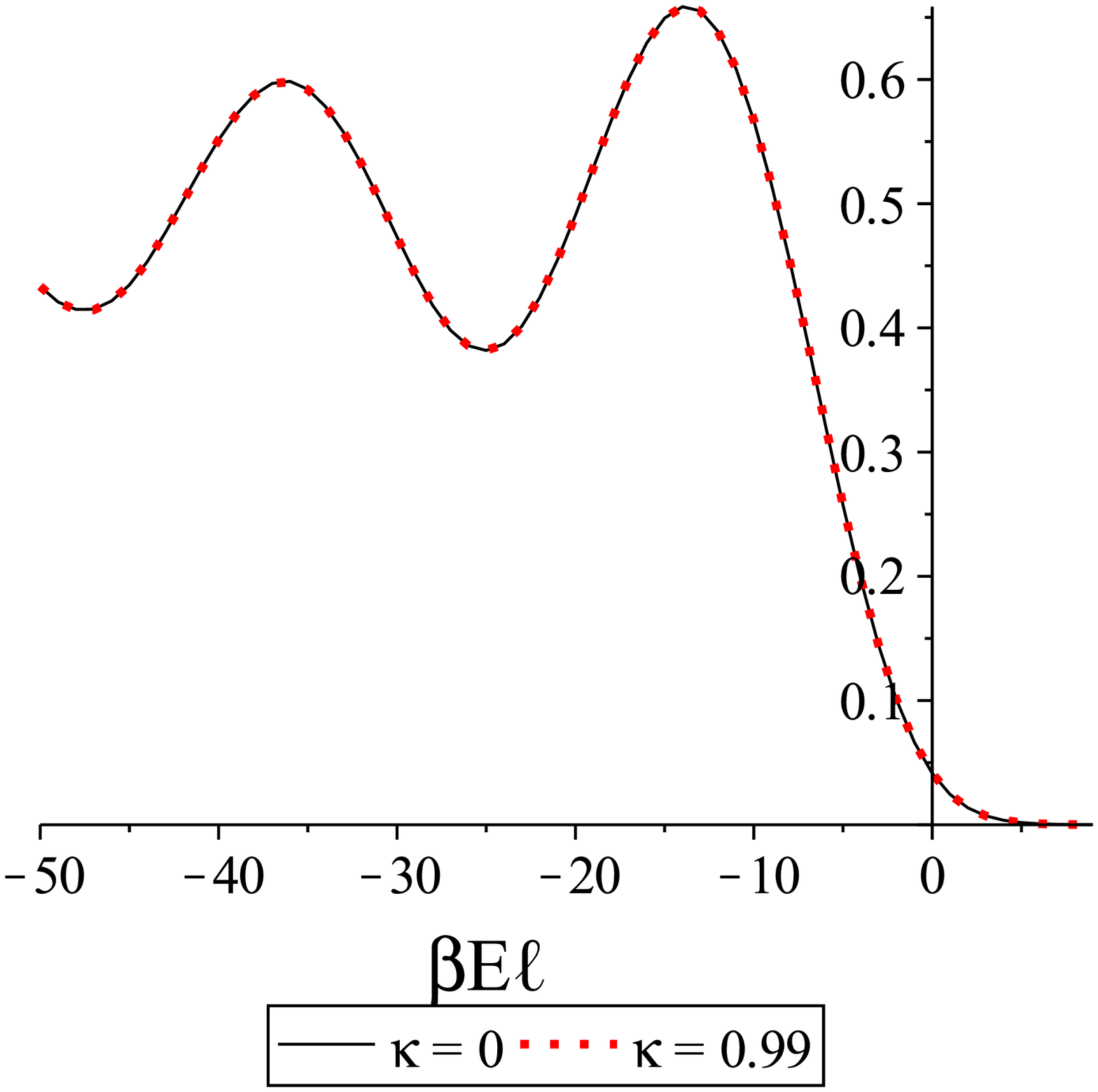}}  
  \subfloat[$\zeta=-1$]{\label{fig:Num_bc-1_rp1_rm0_rm0pt99_a2_k3}\includegraphics[width=0.3\textwidth]{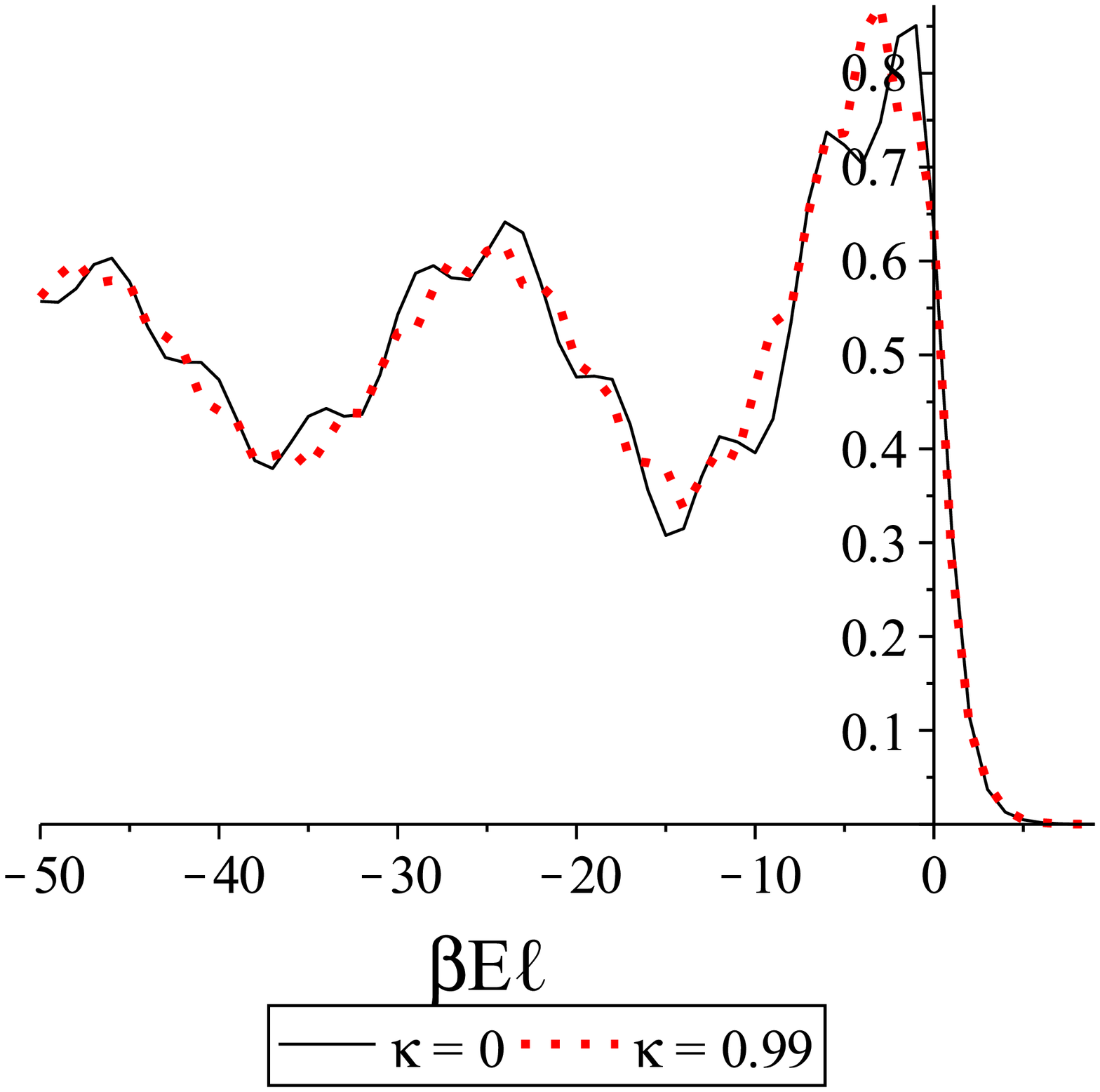}}  
\caption{$\dot{\mathcal{F}}$ as a function of $\beta E\ell$ for 
$r_+/\ell=1$ and $\alpha=2$, with $\kappa =0$ (solid) and $\kappa =0.99$ (dotted) where $\kappa := r_-/r_+$. 
Numerical evaluation from \eqref{eq:corot:rate.evald} with $n\le3$.}
\label{fig:transrate_rp1_a2_rm0_rm0pt99}
\end{figure}

\begin{figure}[t!]
    \centering
  \subfloat[$\zeta=0$]{\label{fig:Num_bc0_rp1_rm0_rm0pt99_a100_k3}\includegraphics[width=0.3\textwidth]{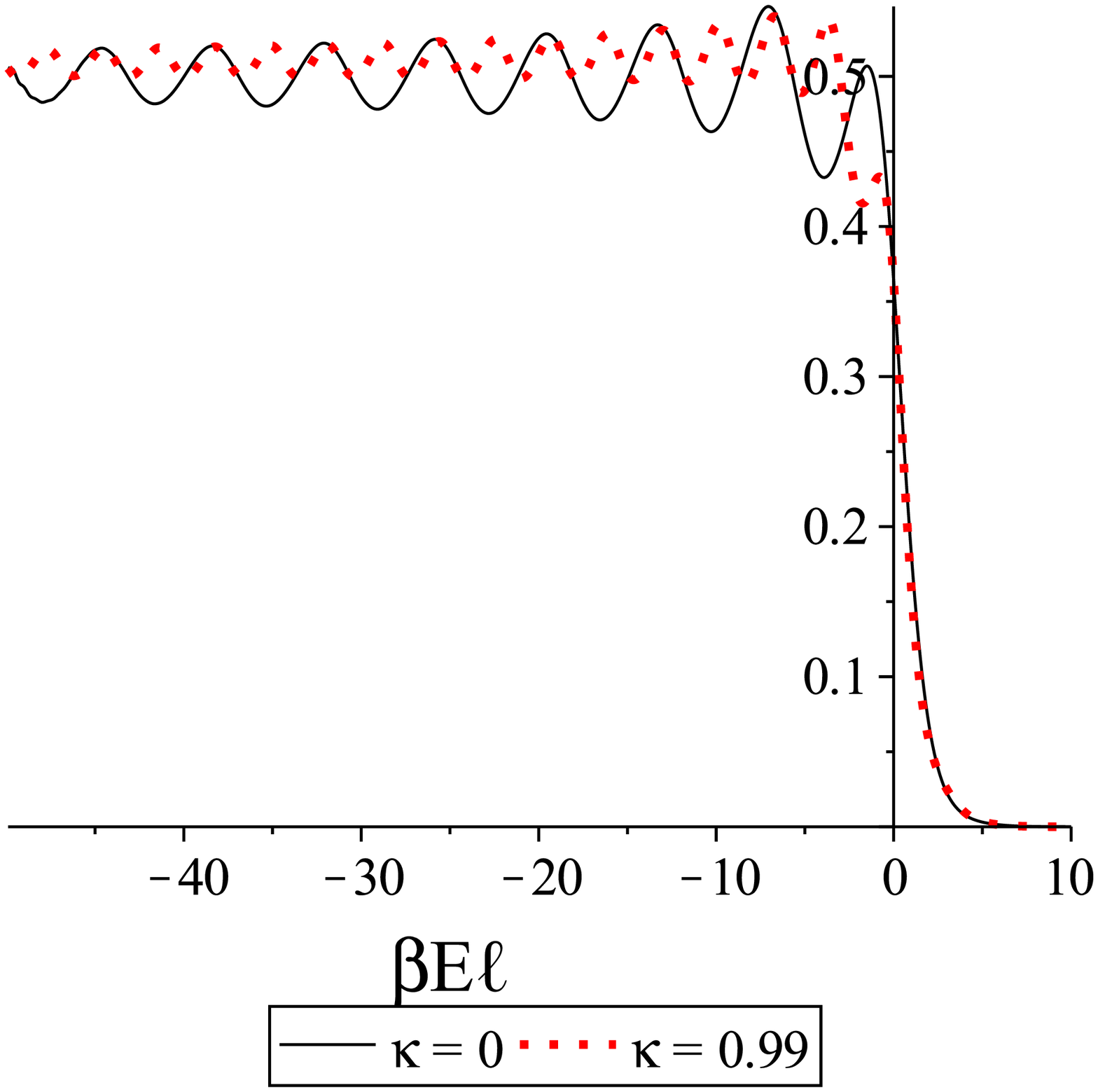}}            
  \subfloat[$\zeta=1$]{\label{fig:Num_bc1_rp1_rm0_rm0pt99_a100_k3}\includegraphics[width=0.3\textwidth]{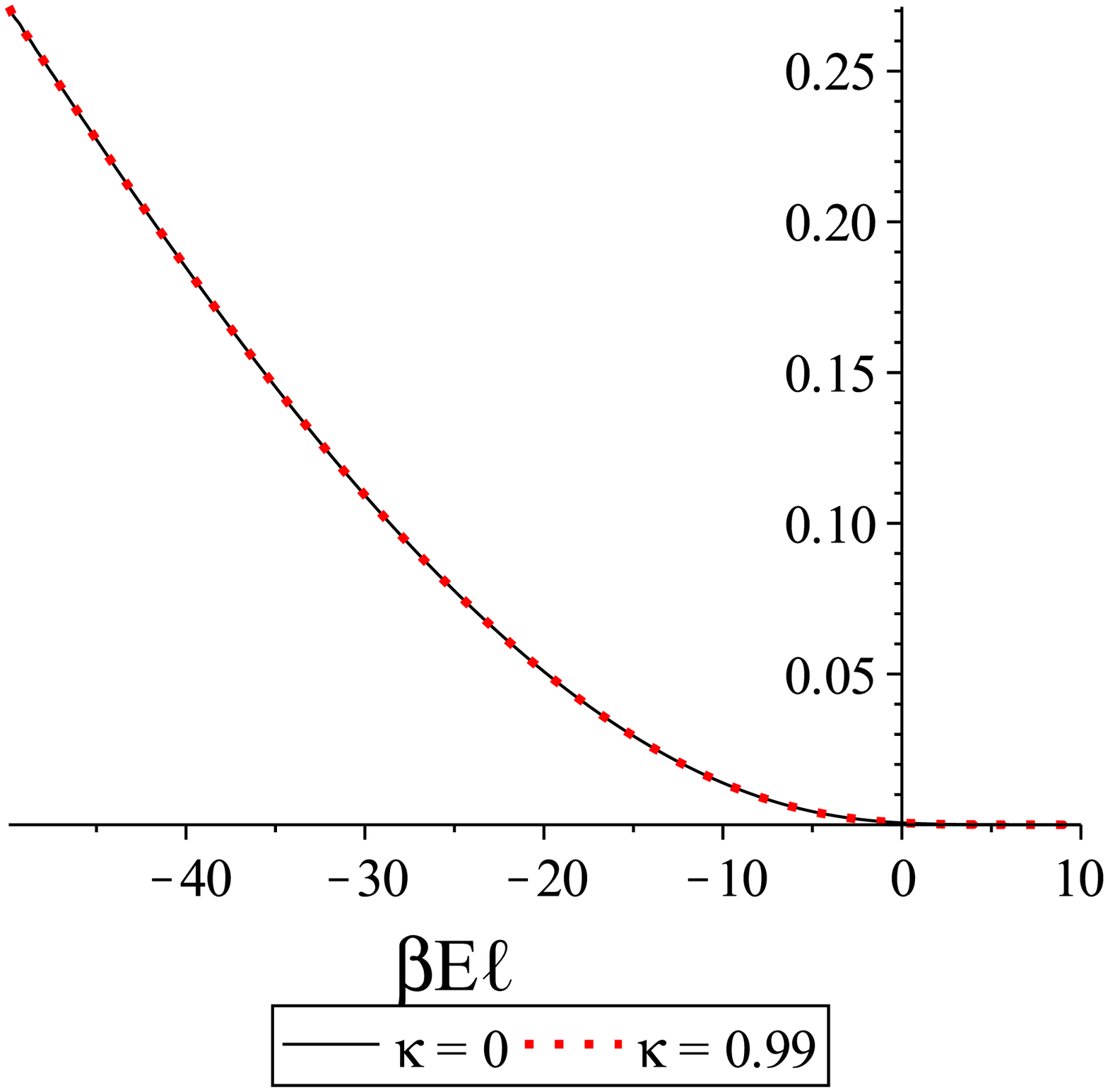}}  
  \subfloat[$\zeta=-1$]{\label{fig:Num_bc-1_rp1_rm0_rm0pt99_a100_k3}\includegraphics[width=0.3\textwidth]{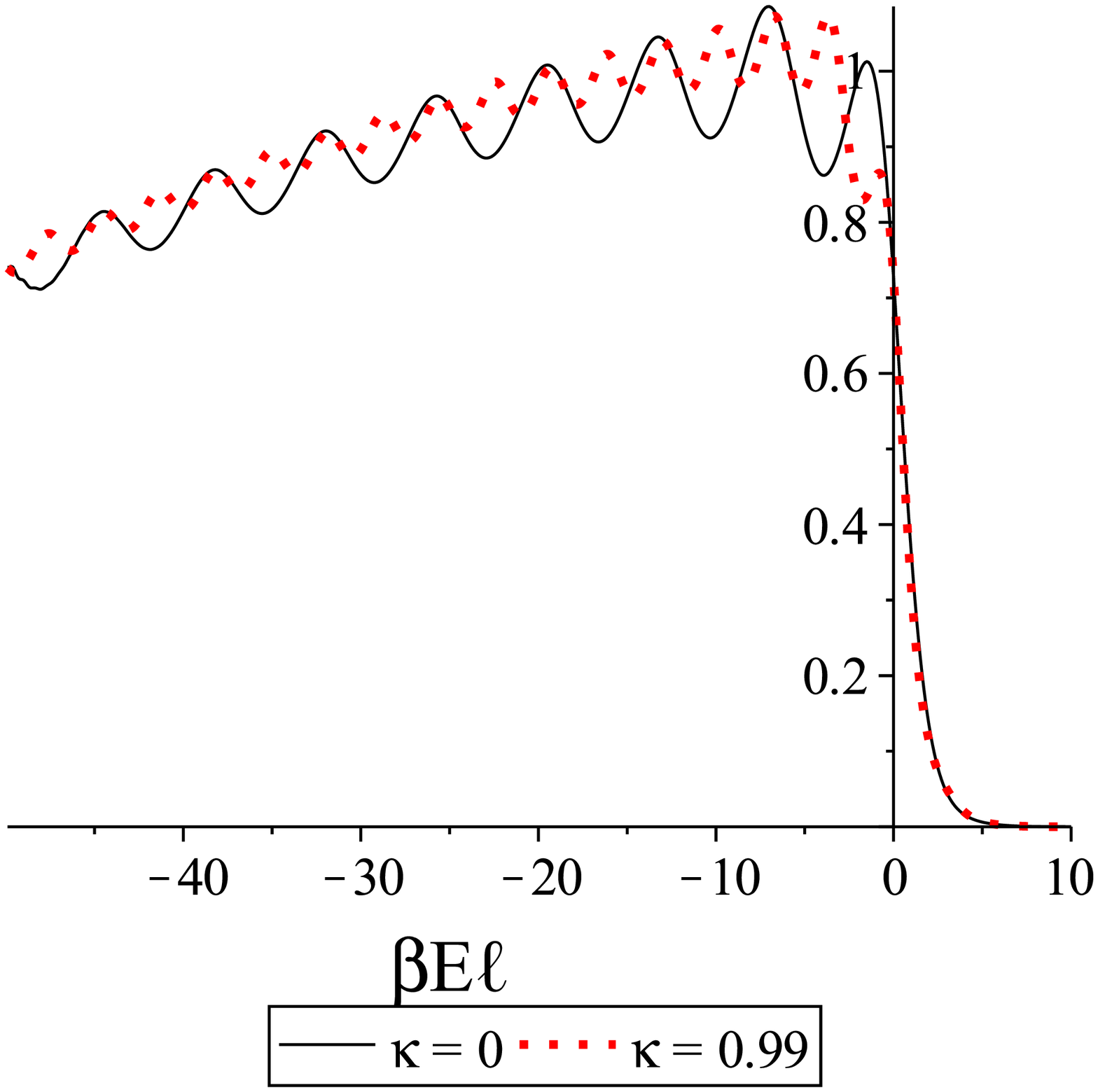}}  
\caption{As in Figure \ref{fig:transrate_rp1_a2_rm0_rm0pt99} but for $\alpha=100$.}
\label{fig:transrate_rp1_a100_rm0_rm0pt99}
\end{figure}

As $r_+/\ell$ decreases, the $n=1$ term in 
\eqref{eq:corot:rate.evald} starts to become appreciable near $r_+/\ell \approx 1$. 
The dependence on $r_-/\ell$ is then no longer exclusively through~$\beta$, and the  
effect is largest for $\zeta=0$ and $\zeta=-1$ but smaller for $\zeta=1$, 
owing to a partial cancellation between the two terms under 
the integral in \eqref{eq:corot:rate.evald} for $\zeta=1$. 
Plots for $r_+/\ell =1$ are shown in Figures
\ref{fig:transrate_rp1_a2_rm0_rm0pt99}
and~\ref{fig:transrate_rp1_a100_rm0_rm0pt99}.

\begin{figure}[p!]
    \centering
  \subfloat[$\zeta=0$]{\label{fig:Num_vs_largeAsy_bc0_rp0pt3_rm0pt299_a2_k3}\includegraphics[width=0.3\textwidth]{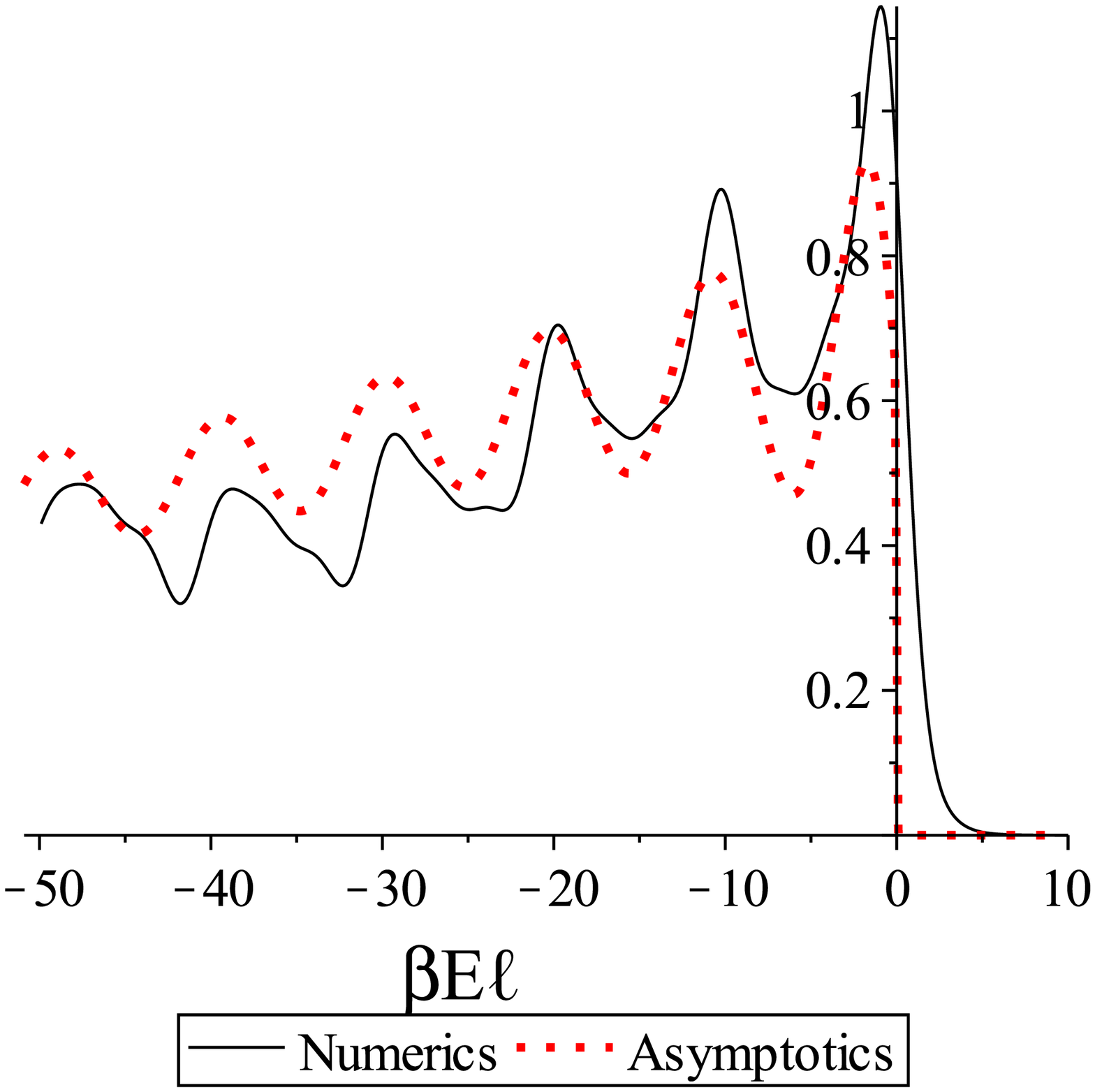}}            
  \subfloat[$\zeta=1$]{\label{fig:Num_vs_largeAsy_bc1_rp0pt3_rm0pt299_a2_k3}\includegraphics[width=0.3\textwidth]{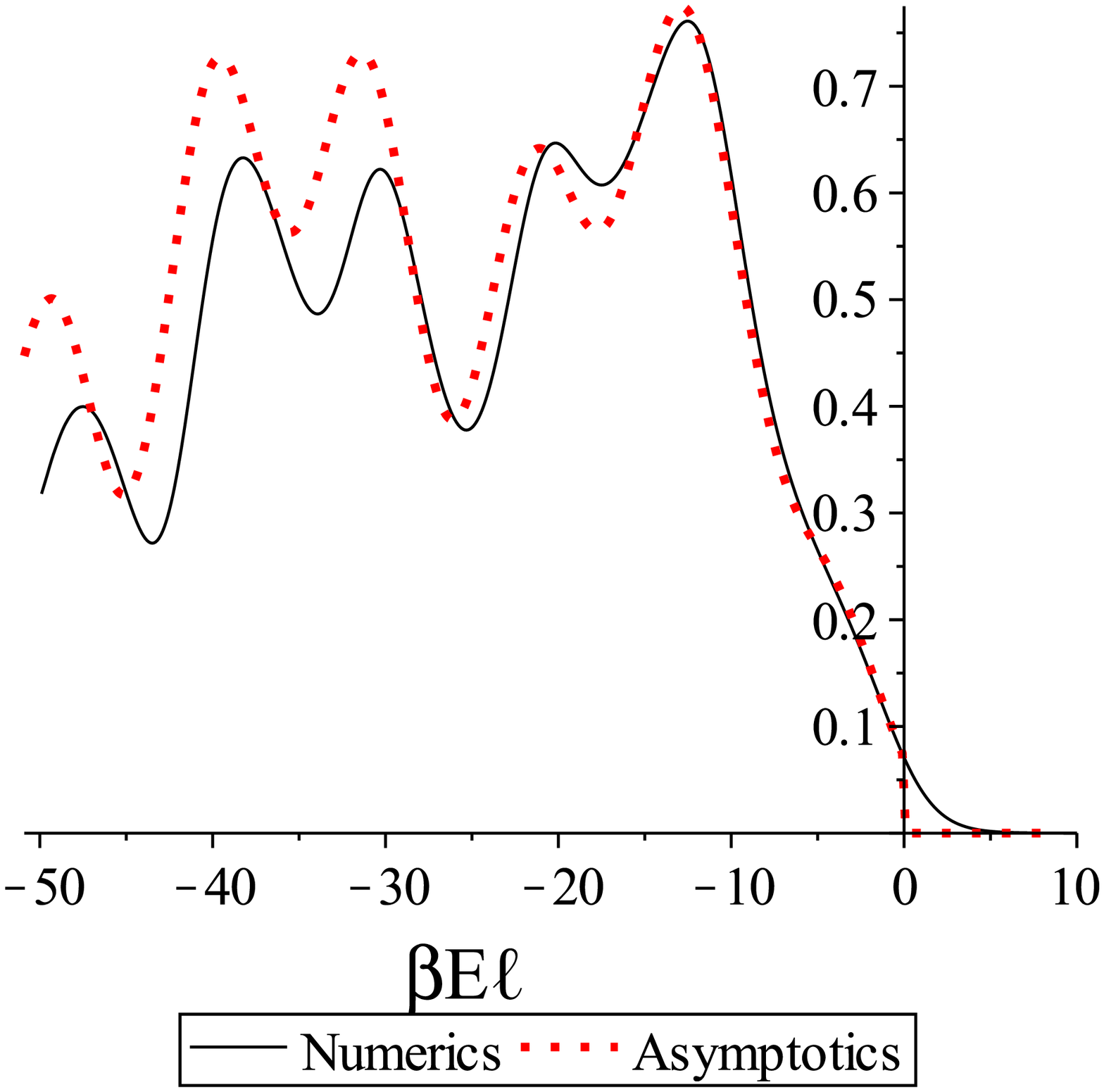}}  
  \subfloat[$\zeta=-1$]{\label{figNum_vs_largeAsy_bc-1_rp0pt3_rm0pt299_a2_k3}\includegraphics[width=0.3\textwidth]{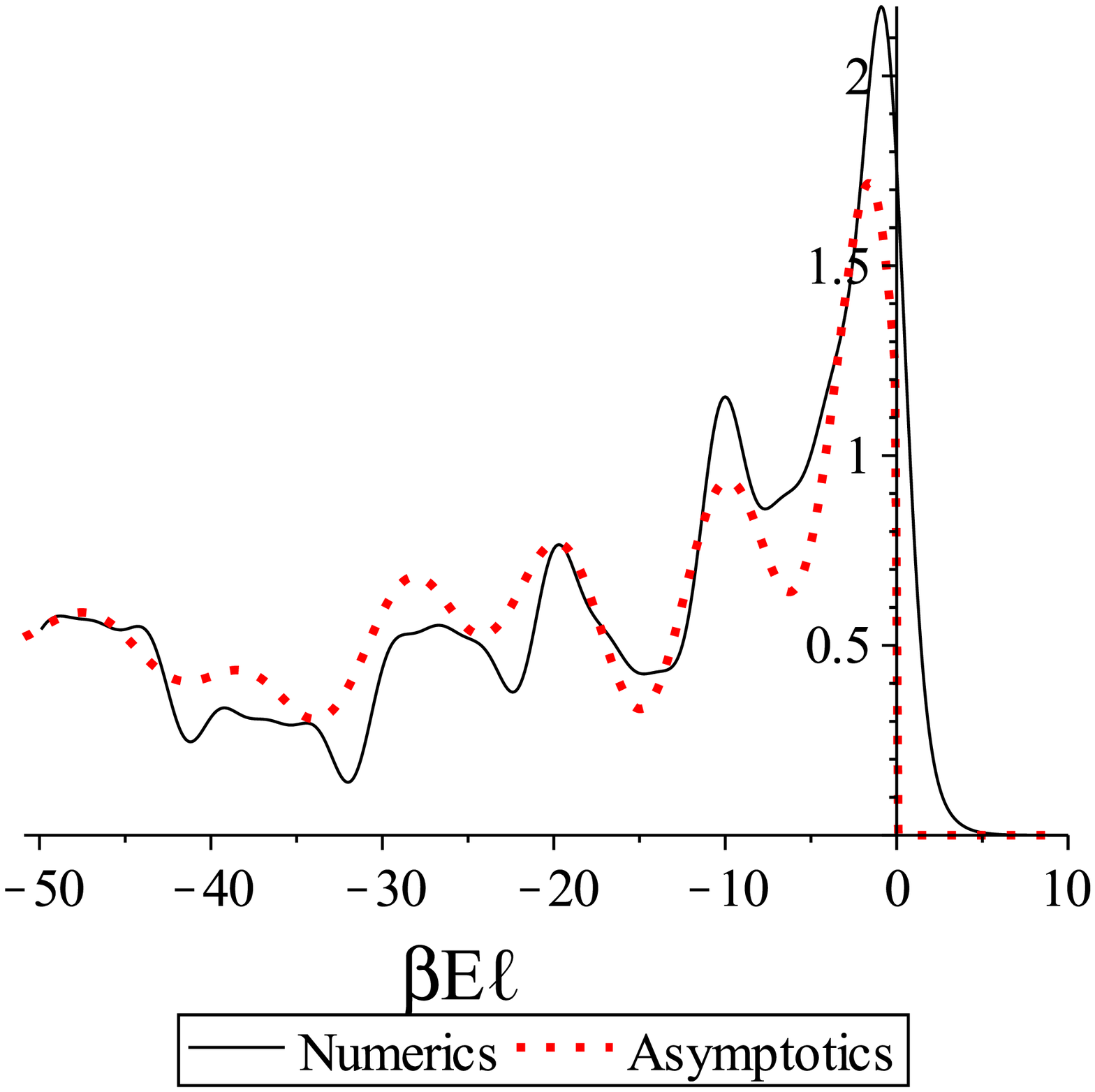}}  
\caption{$\dot{\mathcal{F}}$ as a function of $\beta E\ell$ for 
$r_+/\ell=0.3$ and $r_-/\ell=0.299$, with $\alpha=2$. 
Solid curve shows numerical evaluation from \eqref{eq:corot:rate.evald} with $n\le3$. 
Dotted curve shows the asymptotic large $r_+/\ell$ approximation~\eqref{eq:corot:larger+}.}
\label{fig:Num_vslargeAsy_rp0pt3_a2_rm0pt299}
\end{figure}

\begin{figure}[p!]
    \centering
  \subfloat[$\zeta=0$]{\label{fig:bc0.eps}\includegraphics[width=0.3\textwidth]{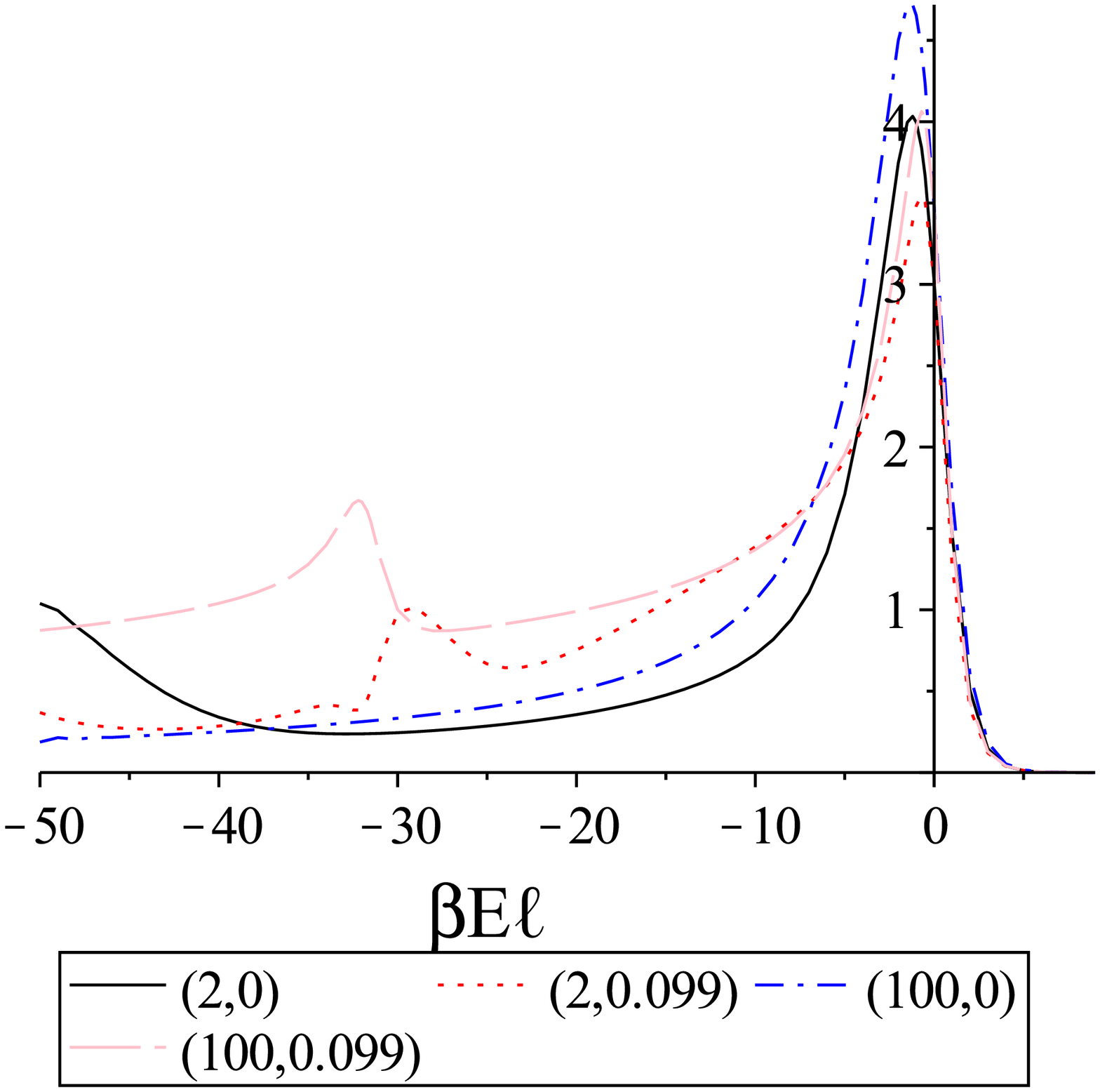}}            
  \subfloat[$\zeta=1$]{\label{fig:Num_bc1_rp0pt01_rm0_rm0pt99_a2_a100_k3}\includegraphics[width=0.3\textwidth]{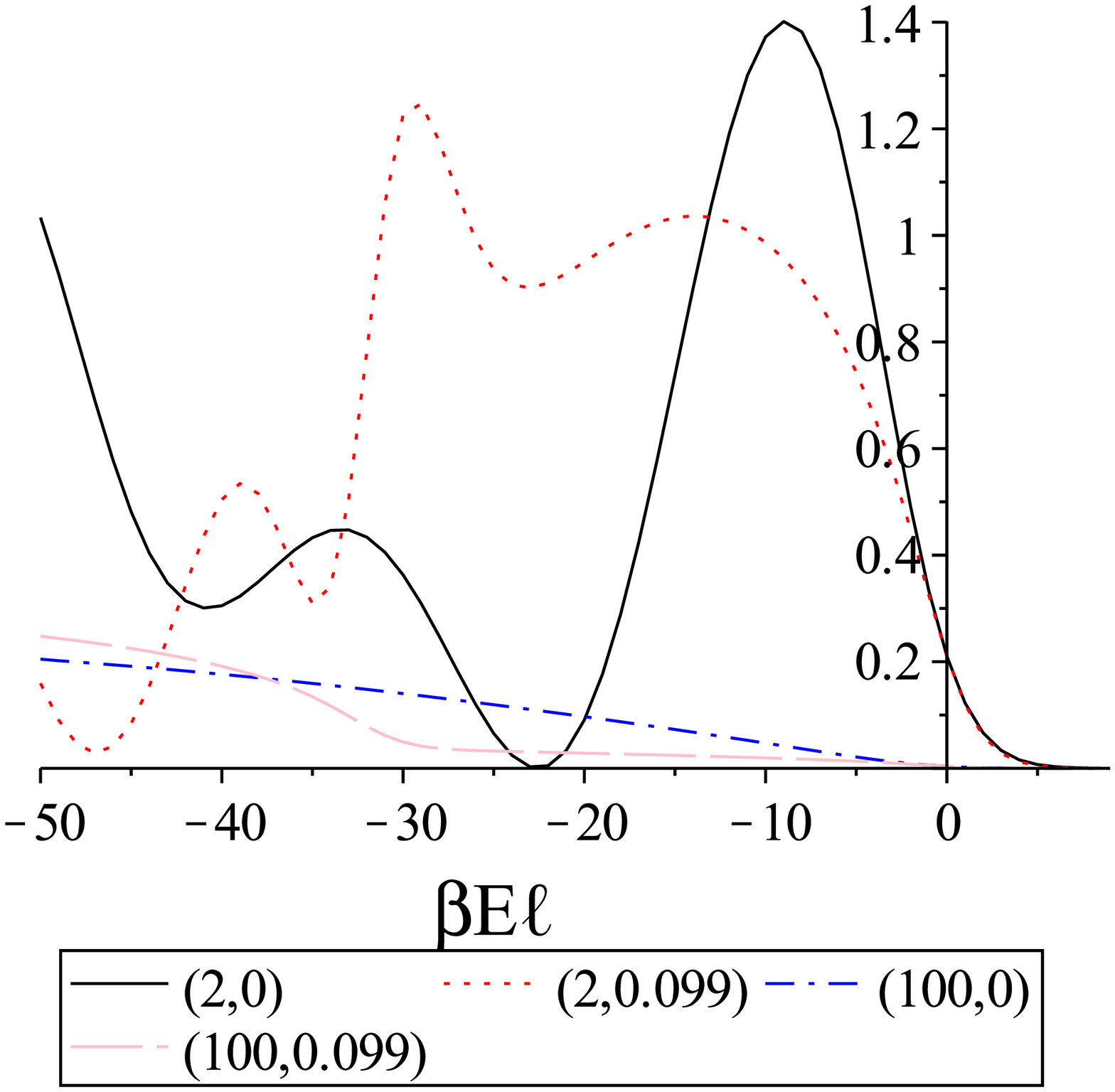}}  
  \subfloat[$\zeta=-1$]{\label{fig:Num_bc-1_rp0pt01_rm0_rm0pt99_a2_a100_k3}\includegraphics[width=0.3\textwidth]{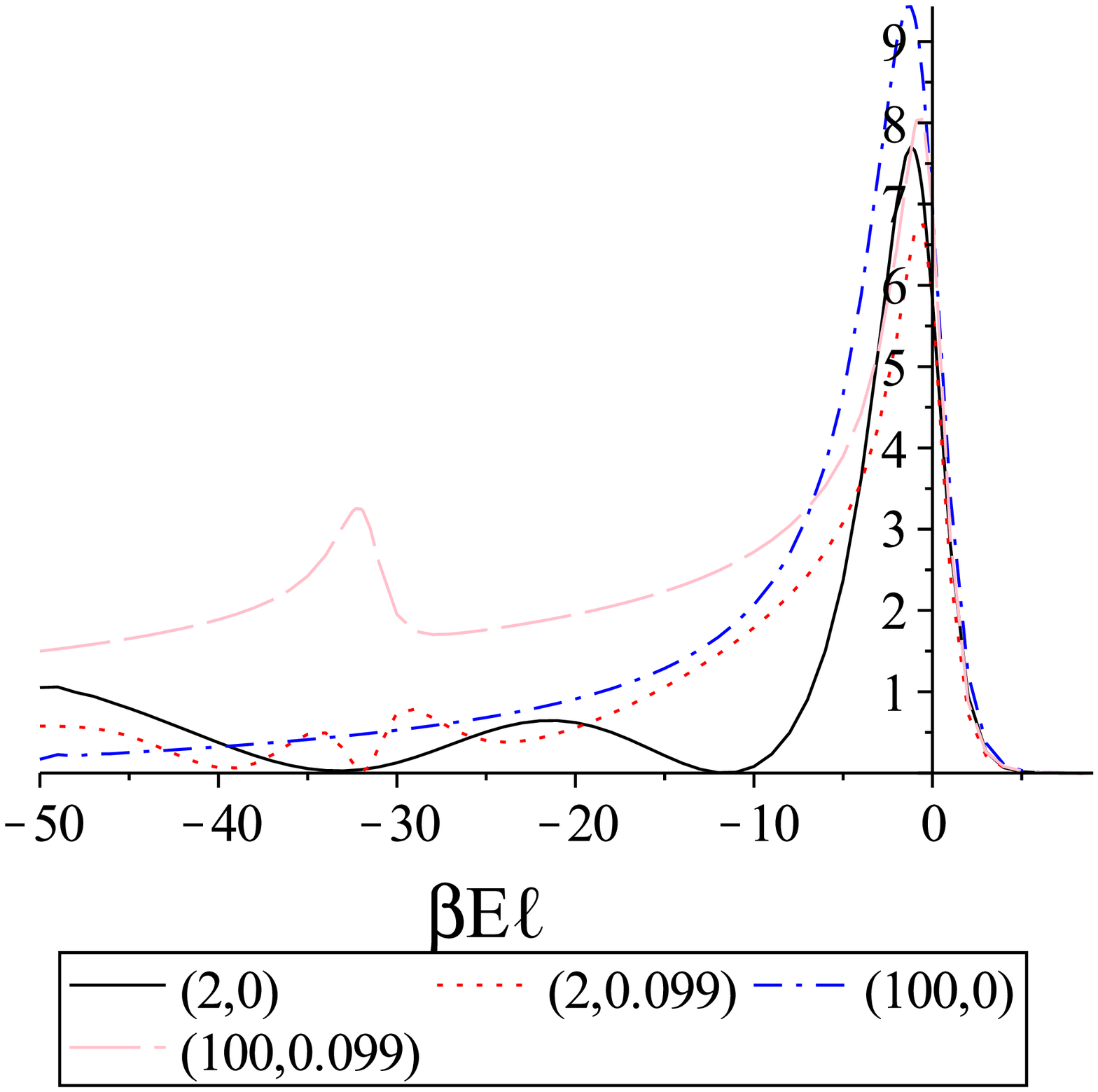}}  
\caption{$\dot{\mathcal{F}}$ as a function of $\beta E\ell$ for 
$r_+/\ell=0.1$, with selected values of the pair $(\alpha,r_-/\ell)$ as shown in the legend. 
Numerical evaluation from \eqref{eq:corot:rate.evald} with $n\le35$.}
\label{fig:transrate_rp0pt1_a2_a100_rm0_rm0pt99}
\end{figure}

As $r_+/\ell$ decreases below~$1$, 
the next-to-leading asymptotic formula \eqref{eq:corot:larger+} 
starts to become inaccurate near $r_+/\ell = 0.3$, 
as shown in Figure~\ref{fig:Num_vslargeAsy_rp0pt3_a2_rm0pt299}, 
although the partial cancellation between the two terms under the integral 
in \eqref{eq:corot:rate.evald} and the similar partial cancellation in 
\eqref{eq:corot:larger+} moderates the effect for $\zeta=1$. 
At $r_+/\ell = 0.1$, shown in Figure \ref{fig:transrate_rp0pt1_a2_a100_rm0_rm0pt99}, 
$\dot{\mathcal{F}}$ is sensitive to changes in both $r_-/\ell$ and~$\alpha$. 
When $\alpha\gg1$, the $\zeta=-1$ curves in Figure 
\ref{fig:transrate_rp0pt1_a2_a100_rm0_rm0pt99} have 
approximately the same profile as the $\zeta=0$ 
curves but at twice the magnitude: from 
\eqref{eq:corot:Kn} and \eqref{eq:corot:Qn} 
we see that this indicates the regime 
where the $n\ge1$ terms in~\eqref{eq:corot:rate.evald} give the 
dominant contribution to~$\dot{\mathcal{F}}$. 

\begin{figure}[t!]
    \centering
  \subfloat[$\zeta=0$]{\label{fig:Num_vs_smallAsy_bc0_rp0pt01_rm0_a4_k300}\includegraphics[width=0.3\textwidth]{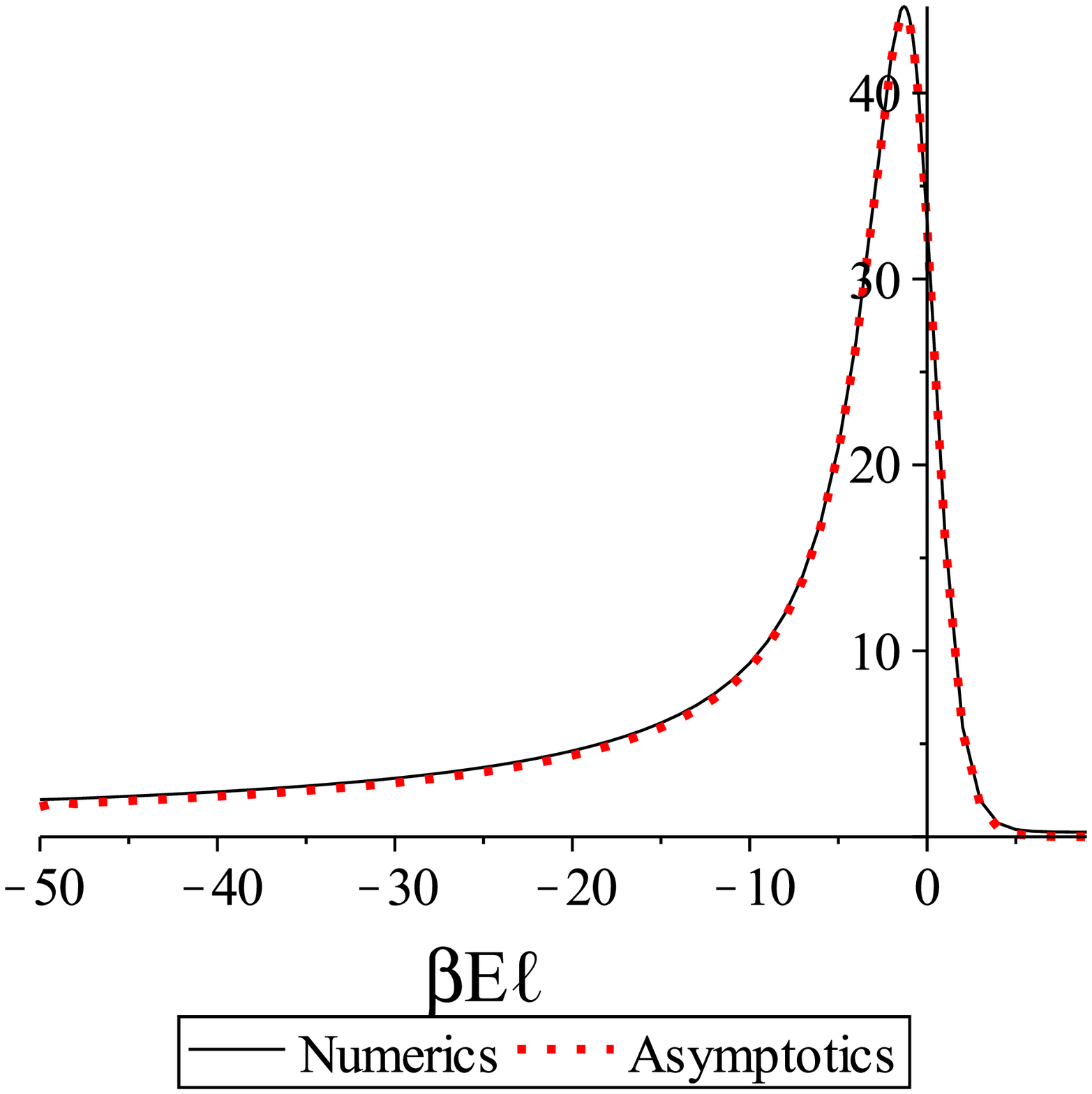}}            
  \subfloat[$\zeta=1$]{\label{fig:Num_vs_smallAsy_bc1_rp0pt01_rm0_a4_k300}\includegraphics[width=0.3\textwidth]{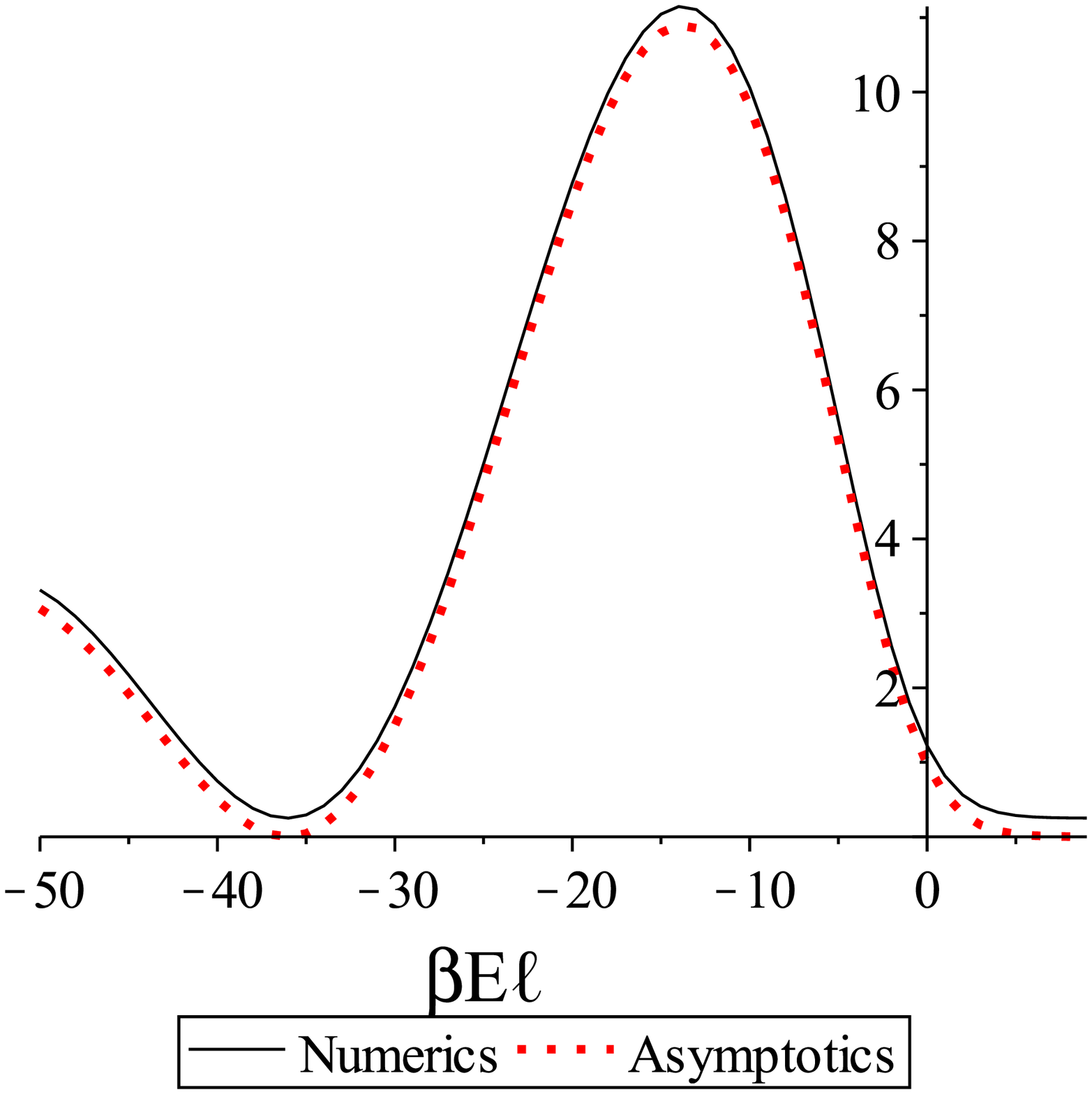}}  
  \subfloat[$\zeta=-1$]{\label{fig:Num_vs_smallAsy_bc-1_rp0pt01_rm0_a4_k300}\includegraphics[width=0.3\textwidth]{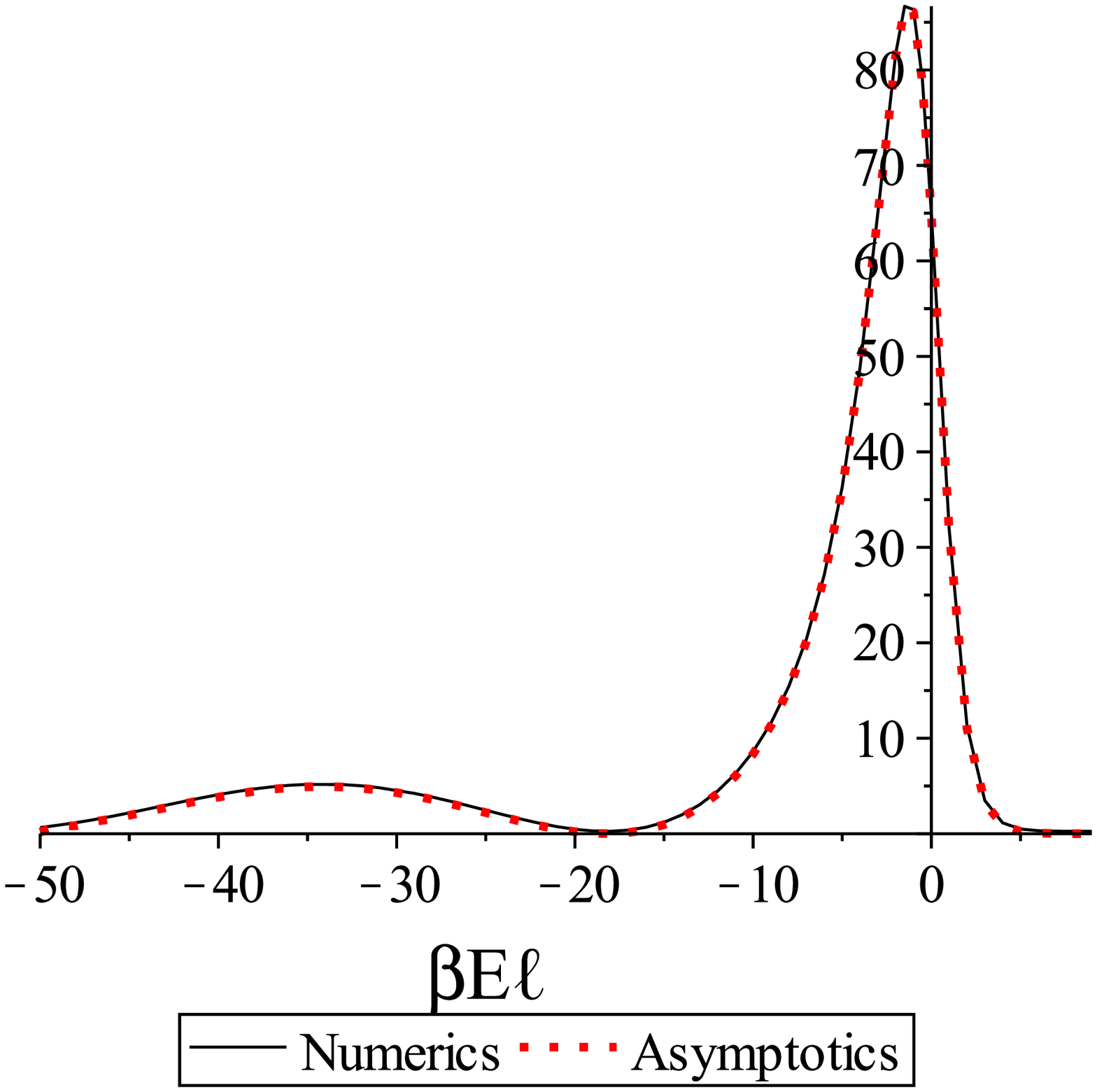}}  
\caption{$\dot{\mathcal{F}}$ as a function of $\beta E\ell$ for 
$r_+/\ell=0.01$ and $r_-=0$, with $\alpha=4$. 
Solid curve shows numerical evaluation from \eqref{eq:corot:rate.evald} with $n\le300$. 
Dotted curve shows the asymptotic small $r_+/\ell$ approximation~\eqref{eq:corot:smallr+}. 
Qualitatively similar graphs ensue for $r_-/r_+=0.99$.}
\label{fig:Num_vs_smallAsy_rp0pt01_rm0_a4_k300}
\end{figure}

As $r_+/\ell$ decreases further, we enter the validity regime of the asymptotic 
formula \eqref{eq:corot:smallr+}, as shown in 
Figure~\ref{fig:Num_vs_smallAsy_rp0pt01_rm0_a4_k300} for $r_+/\ell=0.01$. 
Note that again 
the $\zeta=-1$ curve has approximately the same profile as the $\zeta=0$ 
curve but at twice the magnitude, indicating that the dominant contribution 
comes from the $n\ge1$ terms in~\eqref{eq:corot:rate.evald}. 

\section{Radially in-falling detector in spinless BTZ}
\label{sec:inertial} 

In this section we consider a detector on a radially in-falling
geodesic in a spinless BTZ spacetime. 

\subsection{Transition rate}
\label{sec:inertial:transrate}

Recall from Section
\ref{sec:BTZoverview} that for a spinless hole $r_-=0$ and $r_+ =
M\ell >0$, and the horizon is at $r= r_+$. To begin with, we assume that at least
part of the trajectory is in the exterior region, $r>r_+$. Working in
the exterior BTZ coordinates~\eqref{eq:BoyerLind}, the
radial timelike geodesics take the form 
\begin{align}
&r=\ell\sqrt{M} q \cos\tilde\tau , 
\nonumber\\
&t=\bigl(\ell/\sqrt{M} \, \bigr)
\arctanh\left(\frac{\tan\tilde\tau}{\sqrt{q^2-1}} \right) , 
\nonumber\\
&\phi=\phi_0 , 
\label{eq:inertial:traj}
\end{align}
where $q>1$, $\phi_0$ denotes the
constant value of~$\phi$, and $\tilde\tau$ is an affine parameter such that the proper 
time equals~$\tilde\tau\ell$. 
The additive constants in $\tilde\tau$ and $t$ 
have been chosen so that $r$ reaches its maximum value 
$\ell\sqrt{M}q$ at $\tilde\tau=0$ with $t=0$. 

Substituting \eqref{eq:inertial:traj} in \eqref{eq:BTZ:rate}
and~\eqref{eq:DXN2}, we find that the transition rate is given by 
\begin{align}
&\dot{\mathcal{F}}_{\tau}(E) 
=
1/4
\nonumber\\[1ex]
&\hspace{2ex}
+\frac{1}{2\pi\sqrt{2}}
\sum_{n=-\infty}^{\infty}
\int^{\Delta\tilde\tau}_{0}\,\mathrm{d}\tilde{s}
\Realpart\Biggl[\frac{\mathrm{e}^{-i\Etilde
\tilde{s}}}{\sqrt{-1+K_n\cos\tilde{\tau} \cos( \tilde{\tau}-\tilde{s})
+\sin \tilde{\tau} \sin (\tilde{\tau}-\tilde{s})}}
\nonumber\\[1ex]
&\hspace{19ex}
-\zeta\frac{\mathrm{e}^{-i \Etilde
\tilde{s}}}{\sqrt{1+K_n\cos\tilde{\tau} \cos(\tilde{\tau} -\tilde{s})
+\sin \tilde{\tau} \sin (\tilde{\tau}-\tilde{s})}}
\Biggr], 
\label{eq:inertial:rate}
\end{align}
where 
\begin{equation}
\label{eq:inertial:Kn}
K_n:= 1+2q^2\sinh^2\bigl(n\pi\sqrt{M} \, \bigr) . 
\end{equation}
The detector is switched off at proper time $\tau$ and switched on at proper time 
$\tau_0 = \tau - \Delta\tau$, and we have written 
$\tilde\tau := \tau/\ell$, $\Delta\tilde\tau := \Delta\tau/\ell$
and $\Etilde := E \ell$. 
The square roots in \eqref{eq:inertial:rate} are positive when the
arguments are positive, and they are analytically continued to
negative values of the arguments by giving $\tilde{s}$ a small
negative imaginary part. 

Although the above derivation of \eqref{eq:inertial:rate} proceeded using the exterior BTZ
coordinates, the result \eqref{eq:inertial:rate} holds by analytic
continuation even if the geodesic enters the black or white hole regions. 
The ranges of the parameters are 
$-\pi/2 < \tilde\tau - \Delta\tilde\tau < \tilde\tau < \pi/2$, 
so that the 
detector is switched on after emerging from the white hole singularity 
and switched off before hitting the black hole singularity.

\subsection{The $n=0$ term and KMS}
\label{sec:inertial:zeroth}

We write \eqref{eq:inertial:rate} as 
\begin{align}
\dot{\mathcal{F}}_{\tau} = 
\dot{\mathcal{F}}_{\tau}^{n=0} + \dot{\mathcal{F}}_{\tau}^{n\ne0}
\ , 
\end{align}
where $\dot{\mathcal{F}}_{\tau}^{n=0}$ consists of the $n=0$ term and 
$\dot{\mathcal{F}}_{\tau}^{n\ne0}$ consists of the the sum $\sum_{n\ne0}$. 
We consider first $\dot{\mathcal{F}}_{\tau}^{n=0}$. 

$\dot{\mathcal{F}}_{\tau}^{n=0}$ gives the transition rate of 
a detector on a geodesic in pure~$\text{AdS}_3$. 
$\dot{\mathcal{F}}_{\tau}^{n=0}$
does not depend on $M$ or~$q$, 
and it depends on the switch-on and switch-off moments 
only through~$\Delta\tilde\tau$, that is, through the total detection time. 
Using 
\eqref{eq:inertial:rate} and~\eqref{eq:inertial:Kn}, we find 
\begin{align}
\label{eq:inertial:n0}
\dot{\mathcal{F}}_{\tau}^{n=0}(E) 
=
\frac14
- \frac{1}{4\pi}
\int^{\Delta\tilde{\tau}}_{0}\,\mathrm{d}\tilde{s}\,
\Biggl[\frac{\sin\bigl(\Etilde\tilde{s}\bigr)}{\sin(\tilde{s}/2)}
+\zeta\frac{\cos\bigl(\Etilde\tilde{s}\bigr)}{\cos(\tilde{s}/2)}\Biggr] , 
\end{align}
where $\Etilde := E\ell$. 
As 
$0<\Delta\tilde{\tau}< \pi$, 
\eqref{eq:inertial:n0} is well defined. 

Numerical examination shows that $\dot{\mathcal{F}}_{\tau}^{n=0}$ 
does not satisfy the KMS condition. 
This is compatible with the embedding space
discussion of~\cite{Deser:1997ri,Deser:1998bb,Deser:1998xb,Russo:2008gb}, 
according to which a stationary detector in $\text{AdS}_3$ should respond 
thermally only when its scalar proper 
acceleration exceeds~$1/\ell$. 

The asymptotic behaviour of $\dot{\mathcal{F}}_{\tau}^{n=0}$ 
at large positive and negative energies for fixed $\Delta\tilde{\tau}$
can be found 
by the method of Appendix~\ref{app:zeroterm-largenegenergy}. We find 
\begin{align}
\dot{\mathcal{F}}_{\tau}^{n=0}(E)
&= 
\frac{\Theta\bigl(-\Etilde\bigr)}{2}
+\frac{1}{4\pi \Etilde}
\left(\frac{\cos\bigl(\Etilde\Delta\tilde{\tau}\bigr)}{\sin(\Delta\tilde{\tau}/2)}
-\zeta\frac{\sin\bigl(\Etilde\Delta\tilde{\tau}\bigr)}{\cos(\Delta\tilde{\tau}/2)}\right)
+ O\bigl(1/\Etilde^2\bigr), 
\label{eq:inertial:negE}
\end{align}
where $\Theta$ is the Heaviside step-function.

\subsection{The $n\neq 0$ terms and large $M$ asymptotics}
\label{sec:inertial:nonzero}

We now turn to $\dot{\mathcal{F}}_{\tau}^{n\neq 0}$, which contains the 
dependence of $\dot{\mathcal{F}}_{\tau}$ on $M$ and~$q$. 

We consider $\dot{\mathcal{F}}_{\tau}^{n\neq 0}$ in the limit of large~$M$. 
We introduce a positive constant $c \in (0,\pi/2)$, and we assume that 
the switch-on and switch-off moments are separated from the 
initial and final singularities at least by proper time~$c\ell$. 
In terms of $\tilde\tau$ and $\Delta\tilde\tau$, this means that we assume 
\begin{align}
-\pi/2 + c < \tilde\tau < \pi/2 - c \ , \ 
0< \Delta\tilde\tau < \tilde\tau +\pi/2 - c \ . 
\label{eq:awaysinguniform}
\end{align}

As $K_n=K_{-n}$, we can replace the sum $\sum_{n\ne0}$ in \eqref{eq:inertial:rate} 
by $2\sum_{n=1}^{\infty}$. Given~\eqref{eq:awaysinguniform}, 
the expression 
$\cos\tilde{\tau} \cos( \tilde{\tau}-\tilde{s})$ 
is bounded below by a positive constant. 
Using~\eqref{eq:inertial:Kn}, this implies that the quantities under 
the $n\ne0$ square roots in \eqref{eq:inertial:rate} are
dominated at large $M$ by the term that involves~$K_n$, 
and we may write 
\begin{align}
\label{eq:inertial:LargeM.prep}
\dot{\mathcal{F}}_{\tau}^{n\neq 0} (E)
&=
\frac{1}{\pi\sqrt{2\cos\tilde{\tau}}}
\sum_{n=1}^\infty \frac{1}{\sqrt{K_n}}
\int^{\Delta\tilde{\tau}}_{0}\,\frac{\cos\bigl(\Etilde \tilde{s}\bigr) \, 
\mathrm{d}\tilde{s}}{\sqrt{\cos(\tilde{\tau}-\tilde{s})}}
\left(
\frac{1}{\sqrt{1+ f_-/K_n}}
- 
\frac{\zeta}{\sqrt{1+ f_+/K_n}}
\right)
\end{align}
where 
\begin{align}
f_{\pm} := \frac{\sin \tilde{\tau} \sin (\tilde{\tau}-\tilde{s}) \pm1}{\cos\tilde{\tau} 
\cos(\tilde{\tau} -\tilde{s})} \ . 
\end{align}
The large $M$ expansion of $\dot{\mathcal{F}}_{\tau}^{n\neq 0}$ is then obtained by 
a binomial expansion of the square roots in \eqref{eq:inertial:LargeM.prep} 
at $K_n\to\infty$ and using~\eqref{eq:inertial:Kn}. 
The expansion is uniform in $\tilde\tau$ and $\Delta\tilde\tau$ 
within the range~\eqref{eq:awaysinguniform}, 
and by \eqref{eq:inertial:Kn} it is also uniform in~$q$. 
The first few terms are 
\begin{align}
\dot{\mathcal{F}}_{\tau}^{n\neq 0}(E)
& =
\frac{1}{\pi \sqrt{2\cos\tilde{\tau}}}
\int^{\Delta\tilde{\tau}}_{0}\,\frac{\cos\bigl(\Etilde \tilde{s}\bigr) \, 
\mathrm{d}\tilde{s}}{\sqrt{\cos(\tilde{\tau}-\tilde{s})}}
\Bigg[(1-\zeta) \left(\frac{1}{\sqrt{K_1}} + \frac{1}{\sqrt{K_2}}\right)
+ 
\frac{\zeta f_+ - f_-}{2 K_1^{3/2}}
\Bigg]
\nonumber\\[1ex]
&\hspace{3ex}
+ O \bigl(\mathrm{e}^{-5\pi\sqrt{M}}\bigr)
\ . 
\label{eq:inertial:LargeM.33}
\end{align}
For $\zeta\ne1$, the dominant contribution comes from the term proportional to $(1-\zeta)$ 
and is of order $\mathrm{e}^{-\pi\sqrt{M}}$.

\subsection{Numerical results}
\label{sec:inertial:results}

\begin{figure}[!b]
\centering
\includegraphics[scale=1.5]{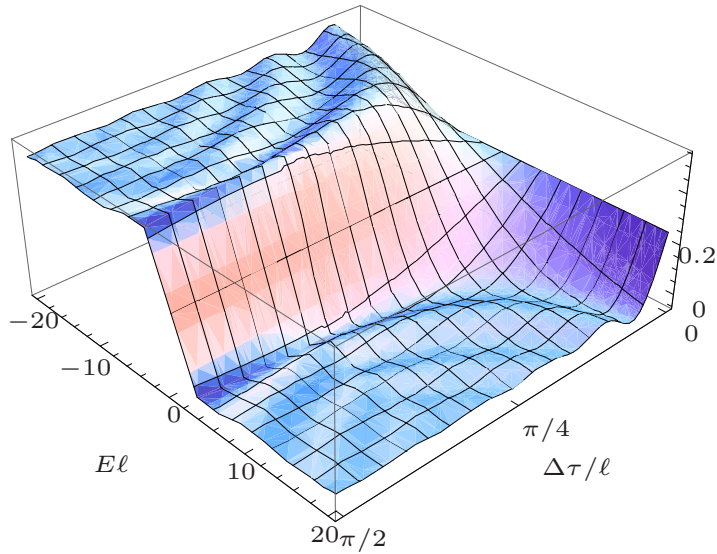}
\caption{$\dot{\mathcal{F}}_{\tau}^{n=0}$ 
\eqref{eq:inertial:n0} as a function of $E\ell$ and 
$\Delta\tau/\ell$ for $\zeta=0$.\label{fig:zeroth3DBC0}}  
\end{figure}

\begin{figure}[!b]
\centering
\subfloat[$E\ell=-100$]{\label{fig:zeroth_E-100}\includegraphics[width=0.32\textwidth]{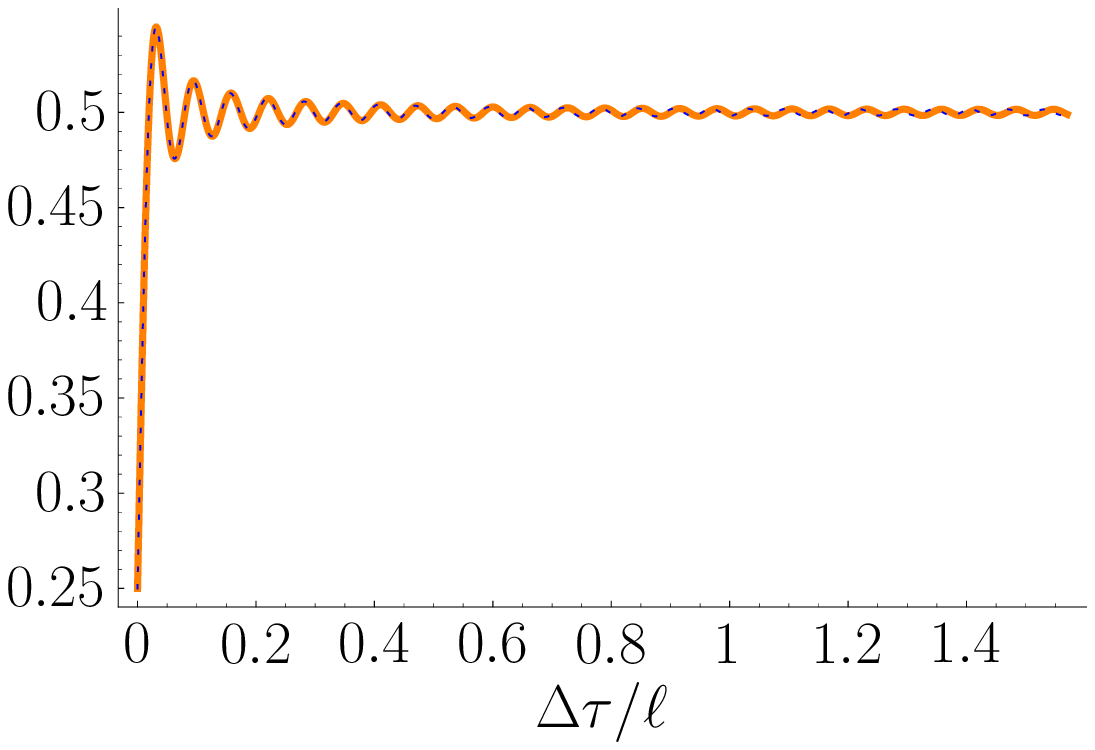}}
\hspace{.1ex}
\subfloat[$E\ell=-5$]{\label{fig:zeroth_E-5}\includegraphics[width=0.32\textwidth]{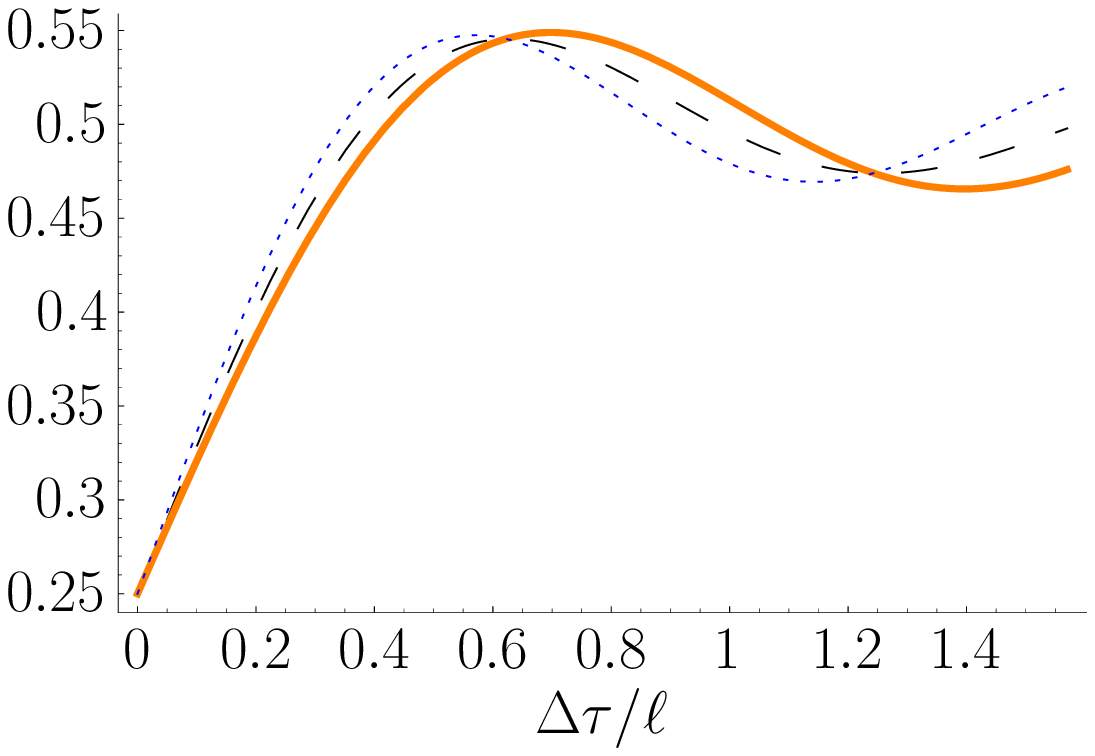}}
\hspace{.1ex}
\subfloat[$E\ell=20$]{\label{fig:zeroth_E20}\includegraphics[width=0.32\textwidth]{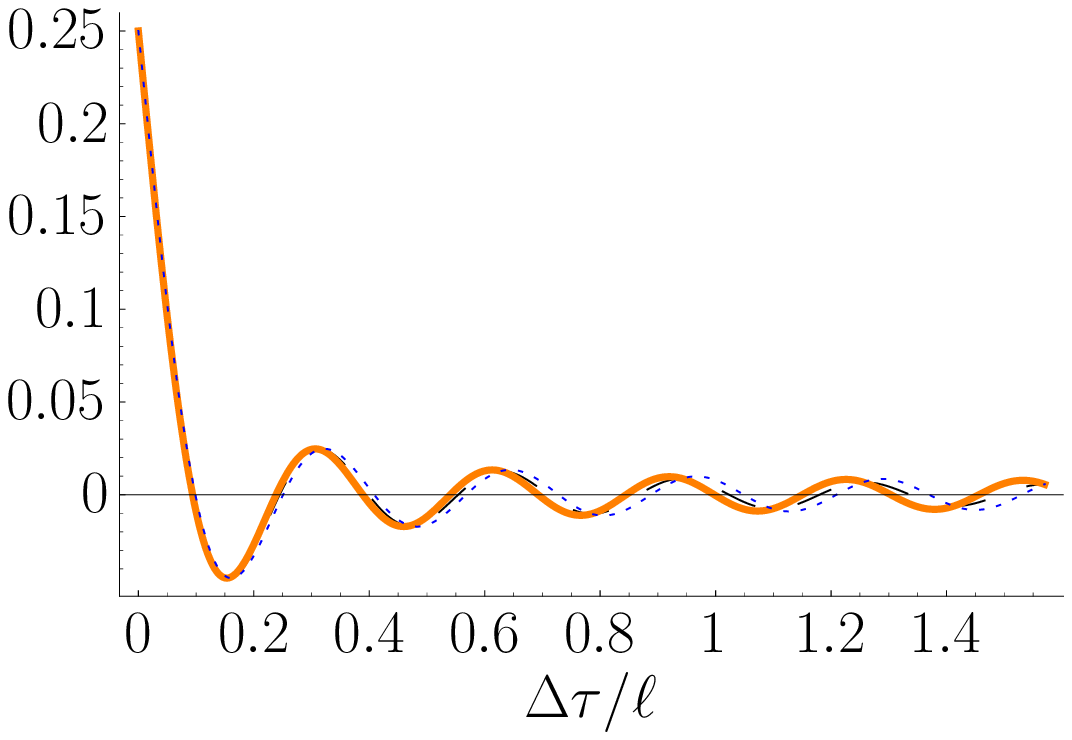}}
\caption{$\dot{\mathcal{F}}_{\tau}^{n=0}$ 
\eqref{eq:inertial:n0} as a function 
$\Delta\tau/\ell$ for selected values of $E\ell$, with
$\zeta=0$ (dashed line), $\zeta=1$ (thick line) 
and $\zeta=-1$ (dotted line).\label{fig:zeroth}}
\end{figure}

At large~$M$, 
the dominant contribution to $\dot{\mathcal{F}}_{\tau}$ 
comes from
$\dot{\mathcal{F}}_{\tau}^{n=0}$~\eqref{eq:inertial:n0}, 
which depends only on $E\ell$ and~$\Delta\tau/\ell$. 
Plots are shown in Figures \ref{fig:zeroth3DBC0} and~\ref{fig:zeroth}.
When $|E\ell|$ is large, the oscillatory dependence on 
$\Delta\tau/\ell$ shown in the plots 
is in agreement with the asymptotic formula~\eqref{eq:inertial:negE}. 

\begin{figure}[t]
\centering
\subfloat[$\zeta=0$]{\label{fig:zerothVSfdot200_E-5_M0pt0001_q100:BC0}\includegraphics[width=0.32\textwidth]{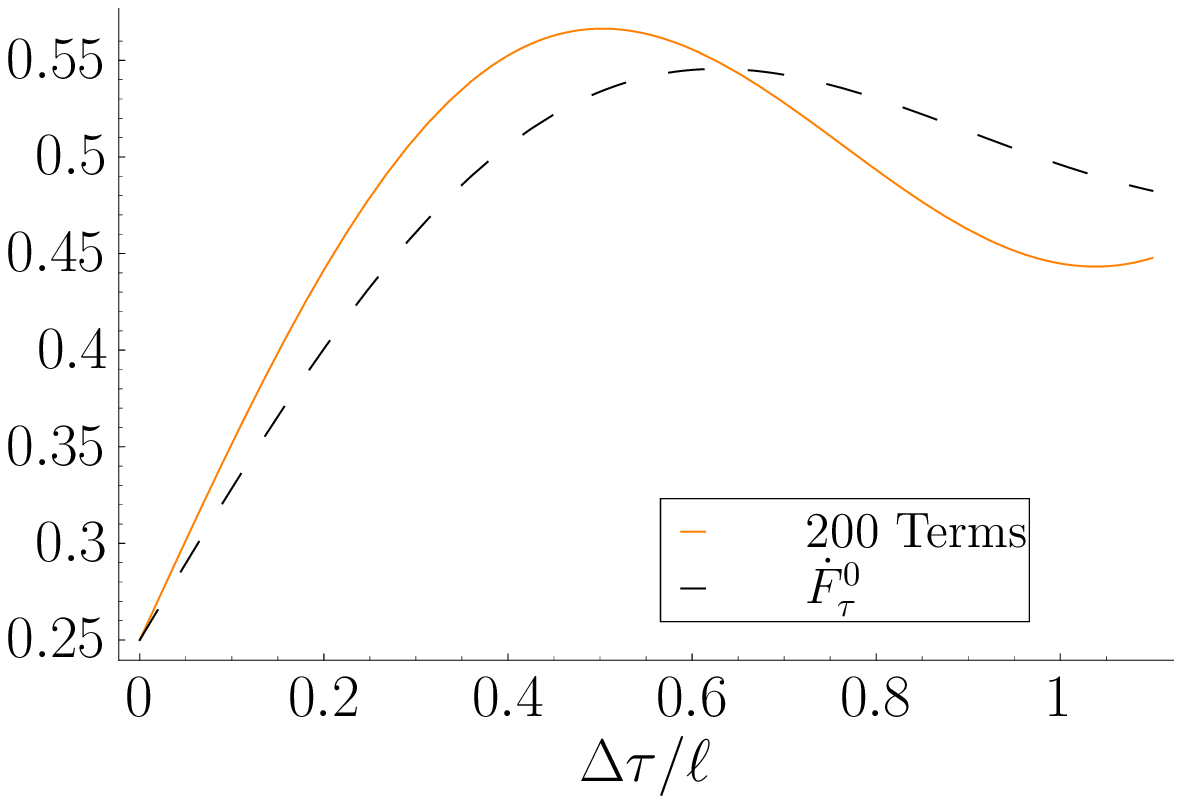}}
\hspace{.1ex}
\subfloat[$\zeta=1$]{\label{fig:zerothVSfdot200_E-5_M0pt0001_q100:BC1}\includegraphics[width=0.32\textwidth]{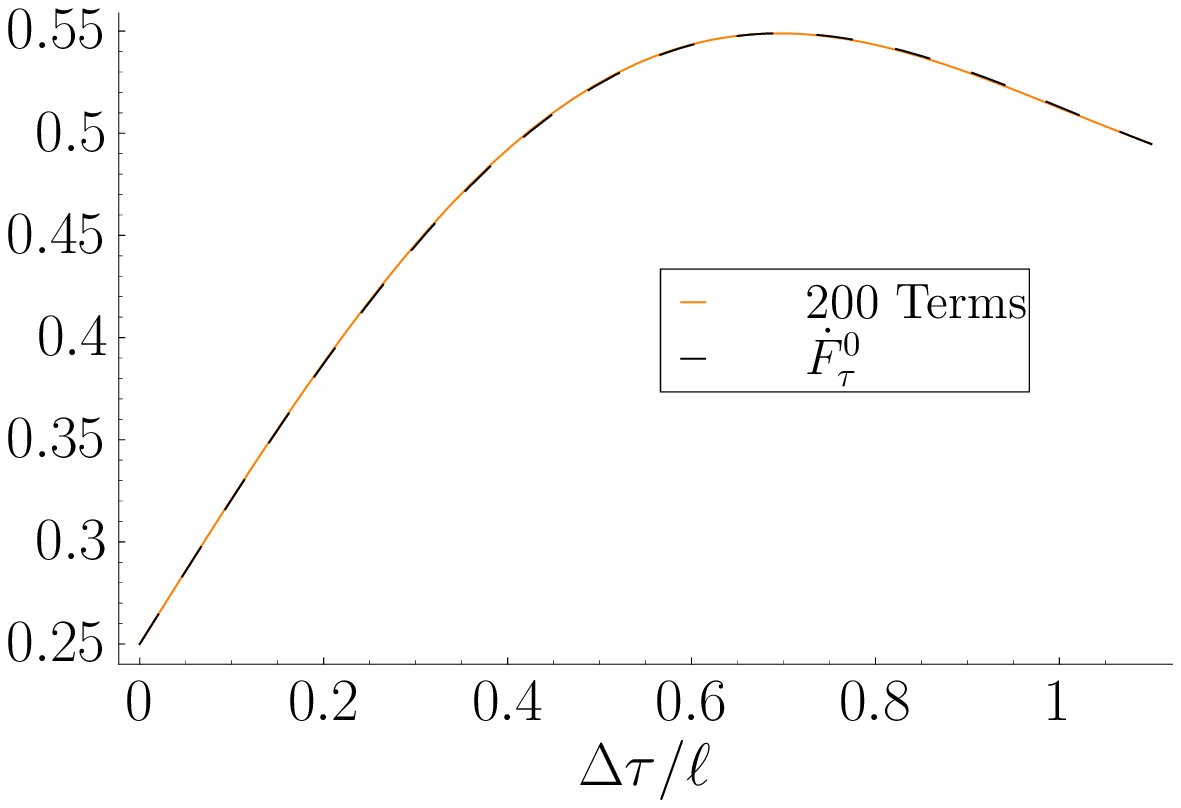}}
\hspace{.1ex}
\subfloat[$\zeta=-1$]{\label{fig:zerothVSfdot200_E-5_M0pt0001_q100:BC-1}\includegraphics[width=0.32\textwidth]{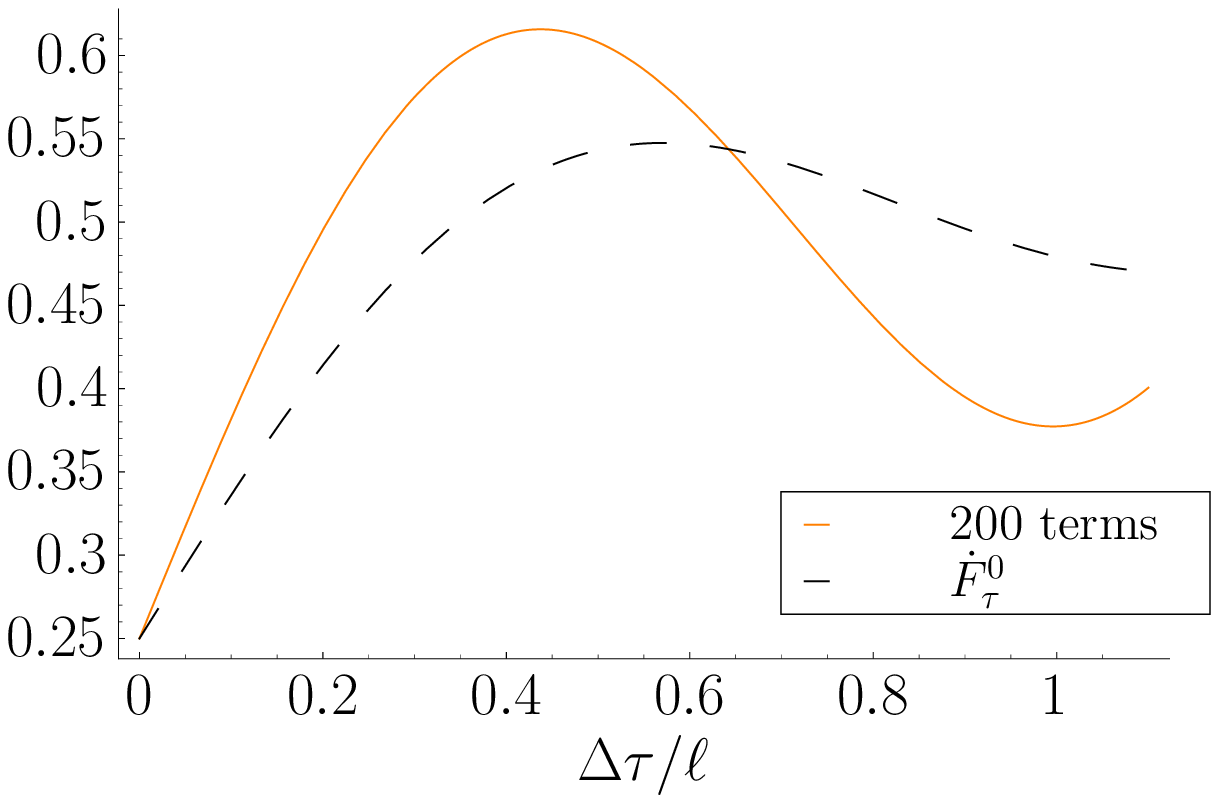}}
\caption{$\dot{\mathcal{F}}_{\tau}$ \eqref{eq:inertial:rate}
with $M=10^{-4}$, 
$q=100$, 
$\tau_0=0$ and $E\ell=-5$. 
Solid curve shows numerical evaluation from \eqref{eq:inertial:rate} 
with $200$ terms 
and dashed curve shows the individual $n=0$ term 
$\dot{\mathcal{F}}_{\tau}^{n=0}$~\eqref{eq:inertial:n0}.
The horizon-crossing occurs outside the plotted range, at 
$\Delta\tau/\ell = \arccos(0.01) \approx 1.56$.\label{fig:zerothVSfdot200_E-5_M0pt0001_q100}}
\end{figure}

\begin{figure}[t]
\centering
\subfloat[$\zeta=0$]{\label{fig:zerothVSfdot200_E20_M0pt0001_q100:BC0}\includegraphics[width=0.32\textwidth]{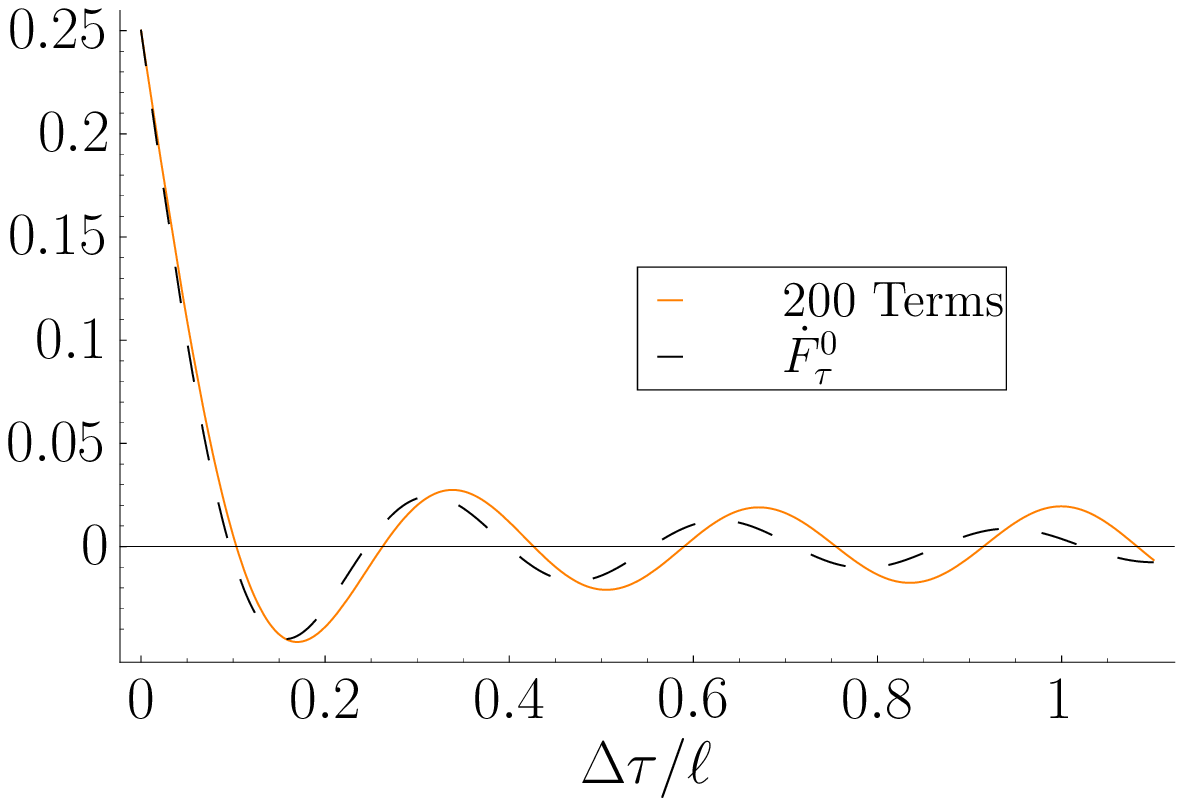}}
\hspace{.1ex}
\subfloat[$\zeta=1$]{\label{fig:zerothVSfdot200_E20_M0pt0001_q100:BC1}\includegraphics[width=0.32\textwidth]{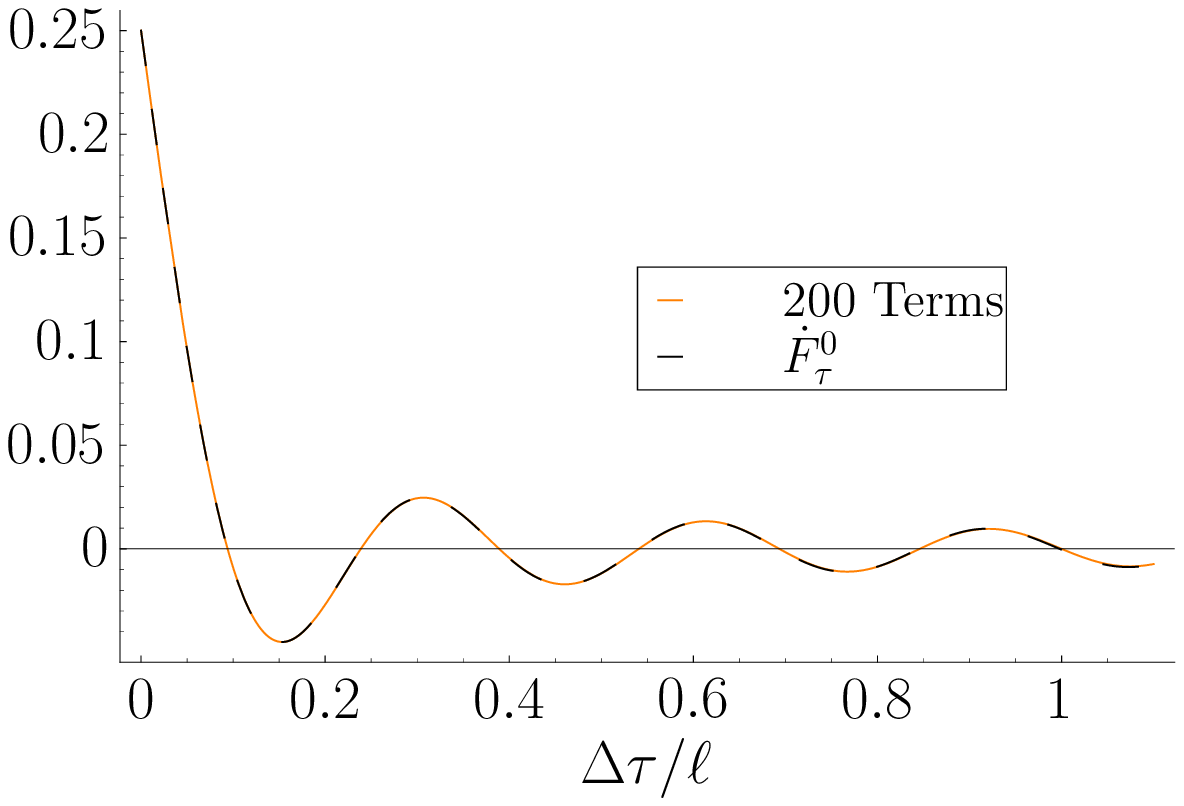}}
\hspace{.1ex}
\subfloat[$\zeta=-1$]{\label{fig:zerothVSfdot200_E20_M0pt0001_q100:BC-1}\includegraphics[width=0.32\textwidth]{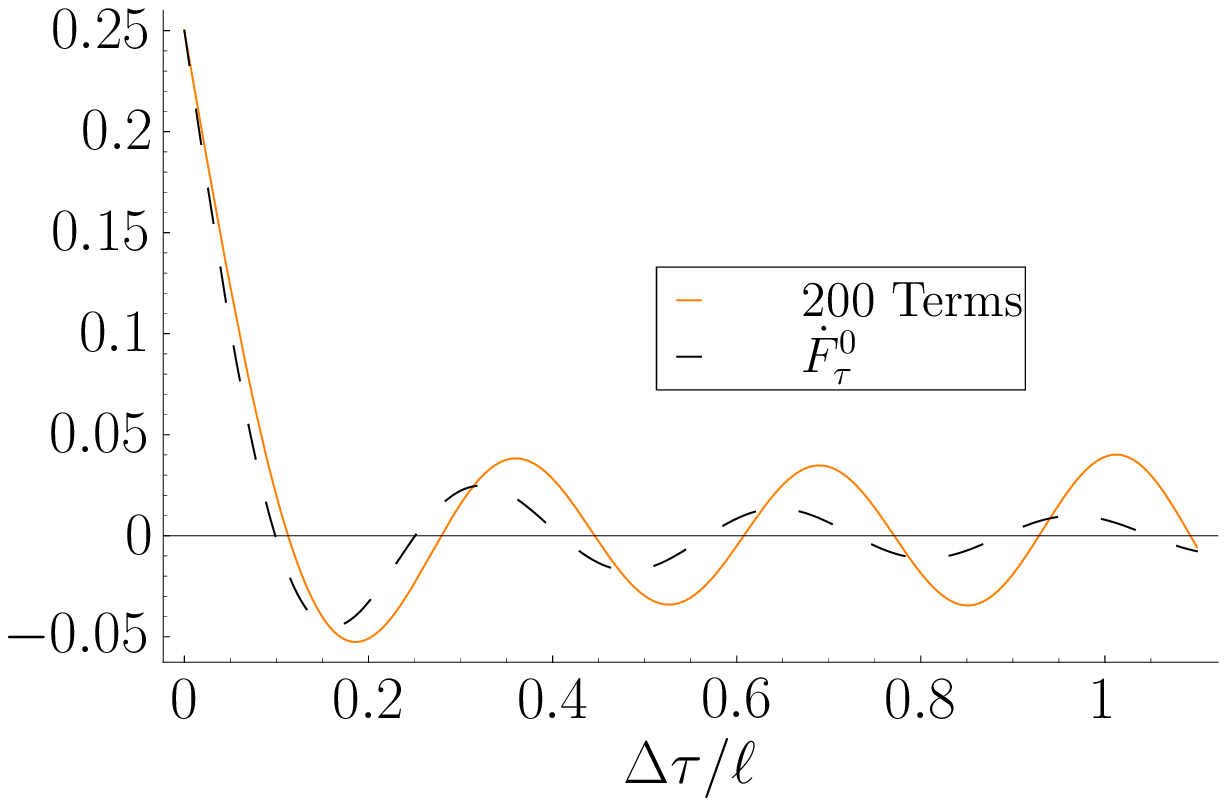}}
\caption{As in Figure \ref{fig:zerothVSfdot200_E-5_M0pt0001_q100} 
but with $E\ell=20$.\label{fig:zerothVSfdot200_E20_M0pt0001_q100}}
\end{figure}

When $M$ decreases, the contribution to 
$\dot{\mathcal{F}}_{\tau}$ from 
$\dot{\mathcal{F}}_{\tau}^{n\ne0}$ becomes significant. For $M=0.1$, 
the terms shown in \eqref{eq:inertial:LargeM.33} are still a good fit to the numerics 
provided both the switch-on and the switch-off are in the exterior region. 
For smaller~$M$, the number of terms that need to be included in 
$\dot{\mathcal{F}}_{\tau}^{n\ne0}$ increases rapidly. 
A~set of plots is shown in Figures 
\ref{fig:zerothVSfdot200_E-5_M0pt0001_q100}
and 
\ref{fig:zerothVSfdot200_E20_M0pt0001_q100}
for $M=10^{-4}$ with $q=100$, taking the detector to be switched on at the moment
where $r$ reaches its maximum and following the detector over a significant fraction 
of its fall towards the horizon. 
$\dot{\mathcal{F}}_{\tau}^{n\ne0}$ turns out to be still insignificant at large negative~$E\ell$, 
but it starts to become significant at $E\ell\gtrsim-5$, and its effect 
then depends strongly on the boundary condition parameter $\zeta$, being the smallest for $\zeta=1$.

For fixed~$M$, following the detector close to the future singularity numerically would pose two complications. First, an increasingly large number of terms would need to be included in $\dot{\mathcal{F}}_{\tau}^{n\ne0}$. Second, the evaluation of the individual terms to sufficient accuracy would need to handle numerically integration over an integrable singularity in~$\tilde{s}$. This singularity arises because the the quantity under the first square root in \eqref{eq:inertial:rate} can change sign within the integration interval. We have not pursued this numerical problem.

\section{Conclusions}
\label{sec:conclusions}

In this paper we have investigated the response of an 
Unruh-DeWitt particle detector in three-dimensional curved spacetime. 
We first obtained a regulator-free expression for the transition probability of a 
detector coupled to a scalar field in an arbitrary Hadamard state, working within first-order perturbation theory and assuming that the detector is switched on and off smoothly, 
and we showed that both the total transition probability and the instantaneous transition rate remain well defined when the switching becomes sharp. In the special case of a detector in Minkowski space, coupled to the Minkowski vacuum of a massless scalar field, these results reduce to those found in~\cite{Hodgkinson:2011pc}. 
The results confirm that the sharp switching limit is qualitatively different between three and four spacetime dimensions even when the spacetime is curved: in four dimensions, the sharp switching limit yields a well-defined transition rate but a divergent transition probability~\cite{satz-louko:curved,satz:smooth}. 

We then specialised to a detector in the BTZ black hole spacetime, coupled to a massless conformally coupled scalar field in a Hartle-Hawking vacuum state with transparent, Dirichlet or Neumann boundary conditions at the infinity. For a stationary detector that is outside the horizon and co-rotating with the hole, and switched on in the asymptotic past, we verified that the transition rate is thermal in the sense of the KMS property, in the local Hawking temperature that is determined by the mass and angular momentum of the hole and by the Tolman redshift factor at the detector's location. 
This is the temperature that was to be expected by general properties of the Hartle-Hawking state 
\cite{Hartle:1976tp,Israel:1976ur,Lifschytz:1993eb,Gibbons:1976ue}, 
and by GEMS considerations 
\cite{Deser:1997ri,Deser:1998bb,Deser:1998xb,Russo:2008gb}. 
A~static detector outside a nonrotating black hole was included as a special case. We obtained analytic results for the transition rate in a number of asymptotic regimes of the parameter space, including those of large and small black hole mass, and we provided numerical results in the interpolating regimes. We have not pursued in detail the case of a stationary detector whose angular velocity differs from that of the hole, but we shall show in Appendix \ref{app:non-corotating} that the parameter space has at least some regimes in which the response of such a detector does not have the KMS property. 

We also considered a detector that falls into a nonrotating BTZ hole along a radial geodesic. As the trajectory is not stationary, the transition rate is not constant along the trajectory, and in particular the switch-on cannot be pushed to the asymptotic past since the trajectory originates at the white hole singularity. We obtained analytic results for the transition rate when the black hole mass is large, 
and we evaluated the transition rate numerically for small values of the black hole mass provided the switch-on and switch-off take place in the exterior. 
We found no parameter ranges where the transition rate would be approximately thermal in the sense of the KMS property, not even near the moment of maximum radius on a trajectory, and we traced the reasons for this to the properties of $\text{AdS}_3$ geodesics that have been previously analysed
from GEMS considerations~\cite{Deser:1997ri,Deser:1998bb,Deser:1998xb,Russo:2008gb}. 
Our expression for the transition rate as a countable sum remains valid after the detector crosses the horizon, but the sum becomes then more difficult to estimate analytically and more labour-intensive to evaluate numerically, and we have not pursued a detailed investigation of this regime. 

It would be interesting to compare our BTZ transition rates to those of a detector on  
similar trajectories in four-dimensional Schwarzschild space-time, 
where the Wightman function needs to be evaluated fully numerically. 
Some differences can be expected to arise from the different 
asymptotic infinities of BTZ and Schwarzschild: 
for example, an 
inertial detector in Schwarzschild should respond to the 
Hartle-Hawking vacuum approximately thermally in the asymptotically flat region. 
We leave this question subject to future work.

\section*{Acknowledgements}
We thank Stephen Creagh for useful discussions on asymptotic techniques 
and Paul Townsend for bringing \cite{Russo:2008gb} to our attention. 
J.~L. thanks the organisers of the ``Bits, Branes, Black Holes'' 
programme for hospitality at the 
Kavli Institute for Theoretical Physics,
University of California at Santa Barbara. 
This research was supported in part by the National 
Science Foundation under Grant No.\ NSF PHY11-25915. 
J.~L. was supported in part by 
the Science and Technology Facilities Council.

\begin{appendix}

\section{Appendix: Derivation of \eqref{eq:corot:rate.evald.alt} and \eqref{eq:corot:rate.evald}}
\label{app:A} 

In this appendix we verify the passage from 
\eqref{eq:corot:rate} to \eqref{eq:corot:rate.evald.alt} 
and~\eqref{eq:corot:rate.evald}. 

\subsection{$n=0$ term}

Let 
\begin{align}
I(a,P) := 
\Realpart
\int_0^\infty \frac{\mathrm{e}^{-iax} \, \mathrm{d}x}{\sqrt{P - \sinh^2 \!x}} \ , 
\label{eq:app:int:Idef}
\end{align}
where $a\in\BbbR$, $P\ge0$, and the square root is 
positive for positive argument and positive imaginary for negative argument. 
We shall show that 
\begin{subequations}
\label{eq:both-singfree}
\begin{align}
I(a,0) &= 
- \frac{\pi \tanh(\pi a/2)}{2}
\ , 
\label{eq:00singfree}
\\[1ex]
I(a,P) &= 
\mathrm{e}^{-\pi a/2}
\int_0^\infty \frac{\cos(ay) \, \mathrm{d}y}{\sqrt{P + \cosh^2 \!y}}
\hspace{5ex} \text{for $P>0$}\ . 
\label{eq:singfree}
\end{align}
\end{subequations}
Applying 
\eqref{eq:both-singfree}
and 
\eqref{eq:00singfree.alt}
to the $n=0$ term in 
\eqref{eq:corot:rate} 
yields the corresponding terms 
in \eqref{eq:corot:rate.evald.alt} 
and~\eqref{eq:corot:rate.evald}. 

Suppose first $P=0$. For $P=0$,  
\eqref{eq:app:int:Idef} reduces to 
$
I(a,0) = 
- \int_0^\infty  
\frac{\sin(ax)}{\sinh x} 
\, \mathrm{d}x 
$, 
which evaluates to \eqref{eq:00singfree}~\cite{gradshteyn}. 

We note in passing the relation 
\begin{align}
I(a,0) = 
-\frac{\pi}{2}
+ 
\mathrm{e}^{-\pi a/2}
\int_0^\infty \frac{\cos(ay) \, \mathrm{d}y}{\cosh y}
\ , 
\label{eq:00singfree.alt}
\end{align}
which follows by 
evaluating the 
integral in \eqref{eq:00singfree.alt} \cite{gradshteyn} and using~\eqref{eq:00singfree}. 
Comparison of 
\eqref{eq:singfree} and \eqref{eq:00singfree.alt} shows that 
$I(a,P)$ is not continuous at $P=0$. 


Suppose then $P>0$. We rewrite \eqref{eq:app:int:Idef} as the contour integral 
\begin{align}
I(a,P) := 
\Realpart \int_{C_1} \frac{\mathrm{e}^{-iaz} \, \mathrm{d}z}{\sqrt{P - \sinh^2 \!z}}
\ , 
\end{align}
where the contour $C_1$ goes from $z=0$ to $z=\infty$ along the positive real axis, 
with a dip in the lower half-plane near the branch point $z=\arcsinh\sqrt{P}$. 
The square root denotes the branch that is positive for small positive~$z$. 


We deform $C_1$ into the union of $C_2$ and~$C_3$, where 
$C_2$ goes from $z=0$ to $z= -i\pi/2$
along the negative imaginary axis 
and 
$C_3$ consists of the half-line 
$z=y-i\pi/2$ with 
$0\le y < \infty$. 
As the integrand has no singularities within the strip $- \pi/2 \le \Imagpart z <0$ 
and falls off exponentially within this strip as $\Realpart z \to +\infty$, 
the deformation does not change the value of the integral. 
The contribution from $C_2$ is purely imaginary and 
vanishes on taking the real part. 
The contribution from $C_3$ yields~\eqref{eq:singfree}. 



\subsection{$n\ne0$ terms}

Let 
\begin{align}
I_b(a,P) := \Realpart 
\int_0^\infty 
\frac{\mathrm{e}^{-iax} \, \mathrm{d}x}{\sqrt{P - \sinh^2 (x+b)}} \ , 
\label{eq:bprotoint}
\end{align}
where $a\in\BbbR$, $P>0$, $b \in \BbbR$ and the square root is 
positive for positive argument and analytically continued to 
negative values of the argument by giving $x$ a small negative imaginary part. 

We shall show that 
\begin{align}
I_b(a,P) + I_{-b}(a,P)
=2\cos{(ab)} \,  I(a,P) \ , 
\label{eq:app:int:pairfinal}
\end{align}
where $I(a,P)$ is given in~\eqref{eq:singfree}. 
Applying 
\eqref{eq:app:int:pairfinal}
with 
\eqref{eq:singfree}
to the $n\ne0$ terms in 
\eqref{eq:corot:rate} 
yields the corresponding terms 
in \eqref{eq:corot:rate.evald.alt}
and~\eqref{eq:corot:rate.evald}. 

For $b=0$, \eqref{eq:app:int:pairfinal} follows from~\eqref{eq:singfree}. 
As both sides of \eqref{eq:app:int:pairfinal} are even in~$b$, it hence suffices to 
consider \eqref{eq:app:int:pairfinal} for $b>0$. 

Let $b>0$. Changing the integration variable in \eqref{eq:bprotoint} to $y = x+b$ yields 
\begin{align}
I_b(a,P) + I_{-b}(a,P)
& =2\cos{(ab)}\Realpart\int^{\infty}_{0} \frac{\mathrm{e}^{-iay} 
\,\mathrm{d}y}{\sqrt{P-\sinh^2 \! y}}
\nonumber
\\[1ex]
& \hspace{1ex}
-\Realpart\left(
\mathrm{e}^{iab}\int^{b}_{0} \frac{\mathrm{e}^{-iay} \,
\mathrm{d}y}{\sqrt{P-\sinh^2 \! y}}
+\mathrm{e}^{-iab}\int^{-b}_{0} \frac{\mathrm{e}^{-iay} \,
\mathrm{d}y}{\sqrt{P-\sinh^2 \! y}}
\right) , 
\label{eq:app:int:pair}
\end{align}
where the branches of the square roots
are as inherited from~\eqref{eq:bprotoint}: 
positive when the argument is positive and continued to negative argument 
by giving $y$ a small negative imaginary part. 
Examination of the branches shows that the last two terms in \eqref{eq:app:int:pair}
cancel on taking the real part, and 
using \eqref{eq:app:int:Idef} in the first term leads to~\eqref{eq:app:int:pairfinal}.

\section{Appendix: Derivation of \eqref{eq:corot:larger+}}
\label{app:B}

In this appendix we verify the asymptotic formula~\eqref{eq:corot:larger+}. 

Let 
\begin{align}
\label{eq:J-def}
J(a,P) := \int_{0}^{\infty} \frac{\cos(ay) \, \mathrm{d}y}{\sqrt{P + \cosh^2 \! y}}
\ , 
\end{align}
where $P>0$ and $a\in\BbbR$. Note from \eqref{eq:singfree} that 
$I(a,P) = \mathrm{e}^{-\pi a/2}J(a,P)$ for $P>0$. 
We shall show that as $P\to\infty$ with fixed~$a$, 
$J(a,P)$ has the asymptotic form 
\begin{subequations}
\begin{align}
J(a,P) &= 
\frac{1}{a \sqrt{\pi P}} 
\Imagpart \bigl[
{(4P)}^{ia/2}
\Gamma(1 + ia/2) \Gamma(\tfrac12 - ia/2) 
\bigr]
+ O\bigl(P^{-3/2}\bigr)
\hspace{3ex}
\text{for $a\ne0$}, 
\label{eq:Jall3}
\\[1ex]
J(0,P) &= 
\frac{1}{2\sqrt{P}} 
\left[
\ln(4P)
+ \psi(1) - \psi(\tfrac12)
\right]
+ O\bigl(P^{-3/2} \ln P\bigr)
\ , 
\label{eq:Jall30}
\end{align}
\end{subequations}
where $\psi$ is the 
digamma function~\cite{dlmf}. 

Starting from~\eqref{eq:J-def}, writing 
$\cos(ay) = \Realpart (\mathrm{e}^{iay})$ and making the substitution 
$y = \ln t$, we find 
\begin{align}
J(a,P) &= 2 \Realpart \int_{1}^{\infty}  
\frac{t^{ia} \, \mathrm{d}t}
{\sqrt{t^4 + B^2 t^2} \, {\displaystyle\sqrt{1 + \frac{1}{t^4 + B^2 t^2}}}}
\nonumber 
\\[1ex]
&= 2 \sum_{p=0}^\infty
b_p \Realpart \int_{1}^{\infty}  
\frac{t^{ia} \, \mathrm{d}t}
{t^{2p+1}{\bigl(t^2 + B^2\bigr)}^{p + (1/2)}}
\ , 
\label{eq:t-powers}
\end{align}
where $B = \sqrt{4 P + 2}$ and 
$b_p$ are the coefficients in the binomial expansion 
${(1+x)}^{-1/2} = \sum_{p=0}^\infty b_p x^p$. 
As the $p>0$ terms in \eqref{eq:t-powers} are 
$O\bigl(B^{-2p - 1}\bigr) = O\bigl(P^{-p - (1/2)}\bigr)$ 
by dominated convergence, we have $J(a,P) = J_0(a,P) + O\bigl(P^{-3/2}\bigr)$, 
where the substitution $t = B v$ in the $p=0$ terms gives 
\begin{align}
J_0(a) &= \frac{2}{B} 
\Realpart \left(
B^{ia}
\int_{1/B}^{\infty}  
\frac{v^{ia-1} \, \mathrm{d}v}
{\sqrt{1+v^2}}
\right) \ . 
\label{eq:Jnought1}
\end{align}

When $a\ne0$, integrating \eqref{eq:Jnought1} by parts and 
extending the lower limit of the integral to zero gives 
\begin{align}
J_0(a,P) &= \frac{2}{Ba} 
\Imagpart \left[
B^{ia}
\int_{0}^{\infty} 
\frac{v^{1+ia} \, dv}
{{\bigl(1+v^2\bigr)}^{3/2}}
+ O\bigl(B^{-2}\bigr)\right]
\ . 
\label{eq:Jnought2}
\end{align}
The integral in \eqref{eq:Jnought2}
may be evaluated by writing 
${(1+v^2)}^{-3/2} = {\bigl(\Gamma(3/2)\bigr)}^{-1}\int_0^\infty
\mathrm{}dx \, x^{1/2} \, e^{-(1+v^2)x}$ and 
interchanging the order of the integrals, 
with the result~\eqref{eq:Jall3}. 
When $a=0$, similar manipulations lead to~\eqref{eq:Jall30}.

\section{Appendix: Derivation of \eqref{eq:corot:smallr+}}
\label{app:C}

Let $p>0$, $q>0$, 
$a\in\BbbR$ and 
$\gamma \in\BbbR$. For $n\in\BbbZ$, let 
$K_n := p^2 \sinh^2 (nq)$, and define 
\begin{align}
F_n := \int_{0}^{\infty} 
\frac{\cos{(n\gamma q)}\cos(ay) \, \mathrm{d}y}{\sqrt{K_n + \cosh^2 \! y}} 
\ , 
\label{eq:Fn}
\end{align}
where we suppress the dependence of $F_n$ on $p$, $q$, 
$a$ and~$\gamma$. We shall show that the sum 
$S := \sum_{n=-\infty}^\infty F_n$
has the asymptotic form 
\begin{align}
S = 
\frac{2}{q}
\int_{0}^{\infty} \mathrm{d}r 
\int_{0}^{\infty} 
\frac{\cos{(r\gamma)}\cos(ay) \, \mathrm{d}y}{\sqrt{p^2 \sinh^2\!r + \cosh^2 \! y}} 
\ + \frac{o(1)}{q}
\label{eq:Fint}
\end{align}
as $q\to0$ with the other parameters fixed. 
Note that the leading term in \eqref{eq:Fint} 
diverges as $q\to0$. 

Let 
\begin{align}
G(r) := \cos(\gamma r) 
\int_{0}^{\infty} 
\frac{\cos(ay) \, \mathrm{d}y}{\sqrt{p^2 \sinh^2\!r + \cosh^2 \! y}} 
\ , 
\label{eq:Gr}
\end{align}
where we suppress the dependence of $G$ on $a$ and~$\gamma$. 
$S$~then equals $q^{-1}$ times the Riemann sum of 
$G$ with the sampling points $r = nq$, $n\in\BbbZ$. 
$G$~is continuous, and from Appendix \ref{app:B} we see that 
$|G(r)|$ is exponentially small as $r\to\pm\infty$. 
The Riemann sum of $G$ therefore converges to the integral of $G$ as $q\to0$.  
Noting finally that $G$ is even, we recover~\eqref{eq:Fint}.

\section{Appendix: Co-rotating response at $E\ell \to\pm\infty$}
\label{app:D}

In this appendix we analyse the individual terms in 
the co-rotating detector response 
\eqref{eq:corot:rate.evald}
in the limit $E\ell\to\pm\infty$. 
These terms are of the form 
\begin{align}
\label{eq:tildeIdef}
\tilde{I}(\chi,a,P):=\cos(\chi a) \, \mathrm{e}^{-\pi a/2} J(a,P)
\ , 
\end{align}
where $\chi\in\BbbR$, $a\in\BbbR$, $P>0$
and $J(a,P)$ is given by \eqref{eq:J-def}. 
We shall show that when $a\to\pm\infty$ with fixed $\chi$ and~$P$, 
$\tilde{I}(\chi,a,P)$ has the asymptotic expansion
\begin{align}
\tilde{I}(\chi,a,P) = 
\begin{cases}
\displaystyle{\frac{2\sqrt\pi \, 
\mathrm{e}^{-\pi a} \cos{(\chi a)}\cos(\alpha a - \pi/4)}{\sqrt{a \sinh(2\alpha)}}
+ o \bigl({a}^{-1/2}\,\mathrm{e}^{-a\pi} \bigr)
\ , }
& a\to+\infty, 
\\[4ex]
\displaystyle{\frac{2\sqrt\pi \cos{(\chi a)}\cos(-\alpha a - \pi/4)}{\sqrt{-a \sinh(2\alpha)}}
+ o \bigl({(-a)}^{-1/2}\bigr)
\ , }
& a\to-\infty, 
\end{cases} 
\label{eq:Itilde-asymptotics}
\end{align}
where $\alpha = \arcsinh\sqrt{P}$. 


Assuming $a\ne0$ and writing $\cos(ay) = \Realpart \bigl(\mathrm{e}^{i|a|y}\bigr)$, we 
start by rewriting 
$J(a,P)$ from \eqref{eq:J-def} as
\begin{align}
\label{eq:JC1}
J(a,P) = \Realpart \int_{C_1} \frac{\mathrm{e}^{i|a|y} \, 
\mathrm{d}y}{\sqrt{P + \cosh^2 \! y}}
\ , 
\end{align}
where the contour $C_1$ consists of the positive imaginary 
axis travelled downwards and the positive real axis travelled rightwards. 
The contribution from the imaginary axis vanishes on taking the real part. 

Writing $P = \sinh^2\!\alpha$ where $\alpha>0$ and 
factorising the quantity under the square root in~\eqref{eq:JC1}, we obtain 
\begin{align}
\label{eq:JC1prime}
J(a,P) = \Realpart \int_{C_1} \frac{\mathrm{e}^{i|a|y} \, \mathrm{d}y}
{\sqrt{\sinh(\alpha+y-i\pi/2) \sinh(\alpha-y+i\pi/2)}}
\ . 
\end{align}
The branch points of the integrand in \eqref{eq:JC1prime} are at 
$y = \pm\alpha + i\pi(n + \frac12)$, $n\in\BbbZ$. 
We may deform $C_1$ into the contour $C_2$ 
that comes down from $\alpha + i\infty$ 
at $\Realpart y =\alpha$, passing the branch points from the left,
encircles the branch point at $y=\alpha+i\pi/2$ counterclockwise, 
and finally goes back up to $\alpha + i\infty$ at $\Realpart y =\alpha$ 
but now passing the branch points from the right. 
Changing the integration variable by 
$y = \alpha + i\pi/2 + iu$, we then have  
\begin{align}
\label{eq:JC2}
J(a,P) = \mathrm{e}^{-|a|\pi/2}
\Realpart 
\left(i \mathrm{e}^{i\alpha |a|}
\int_{C_3} \frac{\mathrm{e}^{-|a|u} \, \mathrm{d}u}
{\sqrt{-i \sin(u) \sinh(2\alpha+iu)}}
\right)
\ , 
\end{align}
where contour $C_3$ comes from 
$u = +\infty$ to $u=0$ on the upper lip of the positive $u$ axis, 
encircles $u=0$ counterclockwise and goes back to $u = +\infty$ on the lower 
lip of the positive $u$ axis. The square root is positive at 
$u = \pi/2$ on the upper lip and 
it is analytically continued to the rest of~$C_3$. 

We now note that $\sinh(2\alpha+iu) = \sinh(2\alpha)\cos(u) + i\cosh(2\alpha)\sin(u)$, 
and that the modulus of this expression is bounded below by~$\sinh(2\alpha)$. 
In~\eqref{eq:JC2}, 
the contribution from the two intervals in which $\pi/2 \le u \le \pi$ is 
therefore bounded above by 
$\mathrm{e}^{-|a|\pi}/\sqrt{\sinh(2\alpha)}$ times a numerical constant, 
and the contribution from the 
two intervals in which $n \pi  \le u \le (n+1)\pi$, $n=1,2,\ldots\,$, 
is bounded above by 
$\mathrm{e}^{-|a|\pi\left[n+(1/2)\right]}/\sqrt{\sinh(2\alpha)}$
times a numerical constant. 
The sum of all of these contributions is hence $O\bigl(\mathrm{e}^{-|a|\pi}\bigr)$. 
In the remaining contribution, coming from the two intervals in which $0 \le u \le \pi/2$, 
we combine the upper and lower lips and change the integration variable to $w = |a|u$. 
This gives 
\begin{align}
\label{eq:C4}
& 
J(a,P)=\frac{2\,\mathrm{e}^{-|a|\pi/2}}{\sqrt{|a|}} \times
\nonumber\\
&\times 
\Realpart\left(\mathrm{e}^{i(\alpha |a|- \pi/4)}
\int_0^{|a|\pi/2} \frac{\mathrm{e}^{-w} \, \mathrm{d}w}
{\sqrt{|a|\sin(w/|a|) 
\bigl[\sinh(2\alpha)\cos(w/|a|) + i\cosh(2\alpha)\sin(w/|a|)\bigr]}}
\right)
\nonumber\\
&
\hspace{2ex}
+ O\bigl(\mathrm{e}^{-|a|\pi}\bigr)
\ , 
\end{align}
where the the square root denotes the branch that is positive in the limit $w\to0_+$. 

By Jordan's lemma, the modulus of the integrand in 
\eqref{eq:C4} 
is bounded from above in the range of integration 
by the function 
$g(w) := \sqrt{\frac{\pi}{2\sinh(2\alpha)}}
\, w^{-1/2} \, e^{-w}$. 
As $g(w)$
is integrable over $0< w < \infty$ and independent of~$a$, 
dominated convergence guarantees that when  
$|a|\to\infty$, the limit in the integrand in \eqref{eq:C4} 
can be taken under the integral. The integral that ensues in the limit is 
elementary, and we obtain 
\begin{align}
J(a,P) = 
\frac{2\sqrt\pi\,\mathrm{e}^{-|a|\pi/2} \cos(\alpha |a|- \pi/4)}
{\sqrt{|a| \sinh(2\alpha)}}
+ o \bigl({|a|}^{-1/2}\,\mathrm{e}^{-|a|\pi/2} \bigr)
\ . 
\label{eq:Ja-as}
\end{align}
\eqref{eq:Itilde-asymptotics} then follows by substituting 
\eqref{eq:Ja-as} in~\eqref{eq:tildeIdef}.


\section{Appendix: Derivation of \eqref{eq:inertial:negE}}
\label{app:zeroterm-largenegenergy}

In this appendix we verify the asymptotic expansions 
\begin{subequations}
\begin{align}
\int^{m}_{0}\,\mathrm{d}x\,\frac{\cos{(\beta x})}{\cos x} &=
\frac{\sin(m\beta)}{\beta\cos m}
+ O\bigl(\beta^{-2}\bigr)
\ , 
\label{eq:app:cosasymp}
\\[1ex]
\int^{m}_{0}\,\mathrm{d}x\,\frac{\sin{(\beta x})}{\sin x}&=
\frac{\pi\sgn\beta}{2}-\frac{\cos{(m\beta)}}{\beta\sin m }
+O\bigl(\beta^{-2}\bigr)
\ , 
\label{eq:app:f-asymp}
\end{align}
\end{subequations}
valid as $\beta\to\pm\infty$ with fixed~$m \in (0,\pi)$. 

\eqref{eq:app:cosasymp} follows by repeated integrations by parts 
that bring down inverse powers of $\beta$~\cite{wong}. 

In \eqref{eq:app:f-asymp}, we split the integral as 
\begin{align}
\label{eq:app:fasymp:C}
\int^m_0\,\mathrm{d}x \left(\frac{1}{\sin{x}} -\frac{1}{x} \right)\sin(\beta x)
- 
\int_m^{\infty}\,\mathrm{d}x\,\frac{\sin{\left(\beta x\right)}}{x} 
+ 
\int_0^\infty\,\mathrm{d}x\,\frac{\sin{\left(\beta x\right)}}{x} 
\ . 
\end{align}
Repeated integrations by parts now apply to the first two terms in~\eqref{eq:app:fasymp:C}, 
and the third term equals $\frac\pi2\sgn\beta$~\cite{gradshteyn}. 
Combining, we obtain~\eqref{eq:app:f-asymp}.

\section{Stationary but non-co-rotating detector}
\label{app:non-corotating}

In this appendix we discuss briefly a detector that is stationary in the 
exterior region of the BTZ black hole but not co-rotating with the horizon. 
For the transparent boundary condition at the infinity, 
we show that the $n=0$ term in the transition rate 
\eqref{eq:BTZ:rate} breaks the KMS property 
already in second order in the difference between the horizon and detector angular velocities. 
As the $n=0$ term is expected to give the dominant contribution when the black hole mass is large, 
we take this as evidence that the transition rate does not satisfy the KMS property, 
in agreement with the GEMS prediction 
\cite{Deser:1997ri,Deser:1998bb,Deser:1998xb,Russo:2008gb}. 

Consider a detector that is stationary in the exterior region of the 
BTZ spacetime at exterior BTZ coordinate~$r$, 
but not necessarily co-rotating with the horizon. 
The tangent vector of the trajectory is a linear 
combination of $\partial_t$ and~$\partial_\phi$. 
By \eqref{eq:BoyerLind} and~\eqref{eq:alpha},  
the lift of the trajectory to $\text{AdS}_3$ reads 
\begin{align}
X_1&=\ell\cosh\chi\sinh(2ky) 
\ , 
\nonumber\\
T_1&=\ell\cosh\chi\cosh(2ky)
\ , 
\nonumber\\
X_2&=\ell\sinh\chi\cosh(2y)
\ , 
\nonumber\\
T_2&=\ell\sinh\chi\sinh(2y)
\ , 
\label{appF:XTXT}
\end{align}
where we have written $\sqrt\alpha = \cosh\chi$ with $\chi>0$, 
the constant $k$ is proportional to the difference 
of the detector and horizon angular velocities, and $y$ is a parameter along the trajectory. 
We assume $|k| < \tanh\chi$, which is the condition for the trajectory to be timelike. 
The proper time $\tau$ is related by $y$ by $\tau = 2 \ell \sinh\chi \sqrt{1 - k^2 \coth^2\!\chi} \, y$. 

Let $\dot{\mathcal{F}}^{n=0}$ denote the $n=0$ term in the transition rate~\eqref{eq:BTZ:rate}. 
Substituting \eqref{appF:XTXT} in~\eqref{eq:DXN2}, and specialising to the 
transparent boundary condition, $\zeta=0$, 
we find 
\begin{align}
\label{eq:transraten=0}
\dot{\mathcal{F}}^{n=0}(E) 
=\frac{1}{4}-\frac{1}{2\pi}\sqrt{1-k^2 \coth^2\!\chi}
\int^{\infty}_{0} \,\mathrm{d}y \, 
\frac{\sin \bigl(2 E\ell  \sinh\chi \, \sqrt{1-k^2 \coth^2\!\chi} \, y\bigr)}
{\sqrt{\sinh^2 \! y-\coth^2\!\chi\sinh^2(k y)}}
\ . 
\end{align}
It can be verified that the quantity under the square root 
in the denominator is positive for $0<y<\infty$. 

Expanding \eqref{eq:transraten=0} as a power series in $E$ and then expanding 
the coefficients as power series in~$k$, we find 
\begin{align}
\dot{\mathcal{F}}^{n=0}(E)
&=\frac{1}{4}
+ 
\left[- \frac{\pi}{4} \sinh\chi + \frac{\pi}{8}\left(\frac{\pi^2}{4}-1\right)
\frac{\cosh^2\!\chi}{\sinh\chi} k^2 + O\bigl(k^4\bigr)
\right]
E\ell
\nonumber
\\[1ex]
& 
\hspace{2ex}
+ \left[
\frac{\pi^3}{12}
\sinh^3\!\chi
+ \frac{\pi^3}{4}
\left(1 - \frac{\pi^2}{6}\right)
\sinh\chi \cosh^2\!\chi 
\, k^2 
+ O\bigl(k^4\bigr)
\right]
{(E\ell)}^3
+ O\bigl({(E\ell)}^5\bigr)
\ . 
\label{eq:transraten=0-expansion}
\end{align}
From \eqref{eq:transraten=0-expansion} it is seen that the
power series expansion of 
$\dot{\mathcal{F}}^{n=0}(-E) / \dot{\mathcal{F}}_{n=0}(E)$ 
in $E$ is incompatible with a pure exponential in~$E$, 
and the discrepancy arises in the coefficient of the ${(E\ell)}^3$ term in order~$k^2$.
$\dot{\mathcal{F}}^{n=0}$ \eqref{eq:transraten=0} hence 
does not satisfy the KMS property at small but nonzero~$k$.

\end{appendix}

\end{document}